\PassOptionsToPackage{unicode}{hyperref}
\PassOptionsToPackage{hyphens}{url}
\PassOptionsToPackage{dvipsnames,svgnames,x11names}{xcolor}
\documentclass[
  12pt]{article}

\usepackage{amsmath,amssymb}
\usepackage{iftex}
\ifPDFTeX
  \usepackage[T1]{fontenc}
  \usepackage[utf8]{inputenc}
  \usepackage{textcomp} 
\else 
  \usepackage{unicode-math}
  \defaultfontfeatures{Scale=MatchLowercase}
  \defaultfontfeatures[\rmfamily]{Ligatures=TeX,Scale=1}
\fi
\usepackage{lmodern}
\ifPDFTeX\else  
\fi
\IfFileExists{upquote.sty}{\usepackage{upquote}}{}
\IfFileExists{microtype.sty}{
  \usepackage[]{microtype}
  \UseMicrotypeSet[protrusion]{basicmath} 
}{}
\makeatletter
\@ifundefined{KOMAClassName}{
  \IfFileExists{parskip.sty}{%
    \usepackage{parskip}
  }{
    \setlength{\parindent}{0pt}
    \setlength{\parskip}{6pt plus 2pt minus 1pt}}
}{
  \KOMAoptions{parskip=half}}
\makeatother
\usepackage{xcolor}
\makeatletter
\ifx\paragraph\undefined\else
  \let\oldparagraph\paragraph
  \renewcommand{\paragraph}{
    \@ifstar
      \xxxParagraphStar
      \xxxParagraphNoStar
  }
  \newcommand{\xxxParagraphStar}[1]{\oldparagraph*{#1}\mbox{}}
  \newcommand{\xxxParagraphNoStar}[1]{\oldparagraph{#1}\mbox{}}
\fi
\ifx\subparagraph\undefined\else
  \let\oldsubparagraph\subparagraph
  \renewcommand{\subparagraph}{
    \@ifstar
      \xxxSubParagraphStar
      \xxxSubParagraphNoStar
  }
  \newcommand{\xxxSubParagraphStar}[1]{\oldsubparagraph*{#1}\mbox{}}
  \newcommand{\xxxSubParagraphNoStar}[1]{\oldsubparagraph{#1}\mbox{}}
\fi
\makeatother

\usepackage{longtable,booktabs,array}
\usepackage{calc} 
\usepackage{etoolbox}
\makeatletter
\patchcmd\longtable{\par}{\if@noskipsec\mbox{}\fi\par}{}{}
\makeatother
\IfFileExists{footnotehyper.sty}{\usepackage{footnotehyper}}{\usepackage{footnote}}
\makesavenoteenv{longtable}
\usepackage{graphicx}
\makeatletter
\def\maxwidth{\ifdim\Gin@nat@width>\linewidth\linewidth\else\Gin@nat@width\fi}
\def\maxheight{\ifdim\Gin@nat@height>\textheight\textheight\else\Gin@nat@height\fi}
\makeatother
\setkeys{Gin}{width=\maxwidth,height=\maxheight,keepaspectratio}
\makeatletter
\def\fps@figure{htbp}
\makeatother

\makeatletter
\@ifpackageloaded{caption}{}{\usepackage{caption}}
\AtBeginDocument{%
\ifdefined\contentsname
  \renewcommand*\contentsname{Table of contents}
\else
  \newcommand\contentsname{Table of contents}
\fi
\ifdefined\listfigurename
  \renewcommand*\listfigurename{List of Figures}
\else
  \newcommand\listfigurename{List of Figures}
\fi
\ifdefined\listtablename
  \renewcommand*\listtablename{List of Tables}
\else
  \newcommand\listtablename{List of Tables}
\fi
\ifdefined\figurename
  \renewcommand*\figurename{Figure}
\else
  \newcommand\figurename{Figure}
\fi
\ifdefined\tablename
  \renewcommand*\tablename{Table}
\else
  \newcommand\tablename{Table}
\fi
}
\@ifpackageloaded{float}{}{\usepackage{float}}
\floatstyle{ruled}
\@ifundefined{c@chapter}{\newfloat{codelisting}{h}{lop}}{\newfloat{codelisting}{h}{lop}[chapter]}
\floatname{codelisting}{Listing}

\makeatother
\makeatletter
\@ifpackageloaded{caption}{}{\usepackage{caption}}
\@ifpackageloaded{subcaption}{}{\usepackage{subcaption}}
\makeatother

\ifLuaTeX
  \usepackage{selnolig}  
\fi
\usepackage[]{natbib}
\usepackage{bookmark}

\IfFileExists{xurl.sty}{\usepackage{xurl}}{} 
\hypersetup{
  pdftitle={Title},
  pdfauthor={Author 1; Author 2},
  pdfkeywords={3 to 6 keywords, that do not appear in the title},
  colorlinks=true,
  linkcolor={blue},
  filecolor={Maroon},
  citecolor={Blue},
  urlcolor={Blue},
  pdfcreator={LaTeX via pandoc}}

\usepackage{multirow}

\newcommand{\anon}{1}

\begin{document}

\def\spacingset#1{\renewcommand{\baselinestretch}%
{#1}\small\normalsize} \spacingset{1}


\if1\anon
{
  \title{\bf Modeling discrete lattice data using the Potts and tapered Potts models}

  \author{Maria Paula Duenas-Herrera\\
    Department of Statistics, The Pennsylvania State University\\
    and \\
    Stephen Berg\\
    Department of Statistics, The Pennsylvania State University\\
    and \\
    Murali Haran\\
    Department of Statistics, The Pennsylvania State University}
  \maketitle
} \fi

\if0\anon
{
  \bigskip
  \bigskip
  \bigskip
  \begin{center}
    {\LARGE\bf Title}
\end{center}
  \medskip
} \fi

\bigskip
\begin{abstract}
The Ising and Potts models, among the most important models in statistical physics, have been used for modeling binary and multinomial data on  lattices in a wide variety of disciplines such as psychology, image analysis, biology, and forestry. However, these models have several well known shortcomings: (i) they can result in poorly fitting models, that is, simulations from fitted models often do not produce realizations that look like the observed data; (ii) phase transitions and the presence of ground states introduce significant challenges for statistical inference, model interpretation, and goodness of fit;  (iii) intractable normalizing constants that are functions of the model parameters  pose serious computational problems for likelihood-based inference.  
      Here we develop a tapered version of the Ising and Potts models that addresses issues (i) and (ii). We develop efficient Markov Chain Monte Carlo Maximum Likelihood Estimation (MCMCMLE) algorithms that address issue (iii). We perform an extensive simulation study for the classical and Tapered Potts models that provide insights regarding the issues generated by the phase transition and ground states. Finally, we offer practical recommendations for modeling and computation based on applications of our approach to simulated data as well as data from the 2021 National Land Cover Database. 

\end{abstract}

\noindent%
\vfill

\newpage
\spacingset{1.8} 

\section{Introduction}\label{sec-intro}

Consider a spatial region that is divided into a collection of subregions or areas, denoted by $D$, and associated with a spatial process $\mathbf{X} = \{X_i : i \in D\}$. When combined with a neighborhood structure that defines how subregions are connected, the process $\mathbf{X}$ forms what is known as a \emph{lattice}. Examples of lattice data include the rates of death from Sudden Infant Death Syndrome (SIDS) in North Carolina counties \citep{Cressie}, satellite-derived land cover type observations over a regular grid \citep{Berrett2012}, and presence or absence data for hermit thrushes in Pennsylvania \citep{Hughes2011}. If each random variable $X_i$ takes values in a finite set of classes $\{1, \ldots, K\}$, we are dealing with multi-categorical data on the lattice. The Ising \citep{lenz1920kernstruktur,ising1924beitrag} and Potts \citep{Potts1952} models have been widely applied to model lattice data in fields as diverse as psychology, image analysis, forestry, and biology, particularly when the goal is 
to analyze the composition of the grid and the degree of spatial correlation in the arrangement. These models, originally proposed in physics, have been used and studied extensively. An informal search on Google Scholar puts the number of citations at over a million, with tens of thousands of citations within just the past five years. Examples of applications of these models include \cite{Liu2010}, where the Ising model was used to analyze how a person's opinion changes according to other people's opinions in a network by studying propagation of influence, and \cite{Berg2019} where the Potts model served as a latent model to label forest communities as part of efforts to describe historical forest composition in Wisconsin. \cite{Murua2012} provides an in-depth explanation of the Potts model for clustering.

While they are popular, the Ising and Potts models have some well known drawbacks that make them difficult to use in applications. First, they often provide a poor fit to observational data when the spatial dependence crosses some threshold. This issue stems from properties intrinsic to their physical origins, known as phase transitions and ground states \citep{Georgii2011}. These properties emerge when the level of spatial correlation exceeds a certain threshold, altering the behavior of the model and potentially compromising its fit to observed data. Second, there are significant computational challenges in inference for these models, primarily posed by the intractable normalizing function present in their  probability mass functions. 

In this work, we study the behavior of the Ising and Potts models under different parameter values using simulations. Based on these simulations, we illustrate the issues that arise from phase transitions and ground states in scenarios characterized by high spatial correlation. We propose a new modeling framework, the \emph{tapered Potts model}, for settings with high spatial correlation. Our approach builds on ideas presented in the context of exponential random graphs \citep{Handcock2017}. We also consider the impact of what is termed an external field; a common practice in Ising and Potts models, usually not justified in applications \citep{Wohrer2019}, is to assume that an external field is absent. We develop algorithms based on Markov chain Monte Carlo maximum likelihood \citep{Geyer1991} to handle the significant computational challenges posed by both the regular and tapered models. 
We find that our approach of introducing a tapering term—(i) addresses the lack of fit of the Potts model in settings with high spatial correlation, (ii) provides flexibility for modeling diverse arrangements of multi-category data on a lattice, and (iii) remains relatively simple to implement. In particular, we offer practical guidance on how to apply this method to real-world data and how to determine when our approach is more appropriate than the classical Potts model.

The remainder of this document is organized as follows. Section \ref{sec-PottsIntro} provides an overview of the classical Ising and Potts models, including some interesting insights on their behavior in the presence of phase transitions and external fields.
We also discuss the challenges this setting presents, along with a heuristic to assess the adequacy of the Potts model. In Section \ref{sec-TapPotts} we introduce the tapered Potts model which we show addresses the problems posed by the standard Potts model. In Section \ref{sec-simStudy} we detail several simulation studies, revealing cases where the tapered Potts model demonstrates improved performance over the traditional Potts model. Section \ref{sec-NLCDIntro} applies these models to subsets of the 2021 National Land Cover Dataset, and Section \ref{sec-Discussion} concludes with a summary and discussion of the findings.

\section{The Ising and Potts models}\label{sec-PottsIntro}

The Potts model, which includes the Ising model as a special case, was originally introduced to describe magnetic behavior in physics. In a statistical context, the Potts model can be viewed as a framework for modeling multi-category data on a lattice by defining a neighborhood structure among lattice cells and assuming that each cell's realization is conditionally independent given the realizations of its neighbors. The Ising model corresponds to the special case of binary data, with just two categories. 

Consider a lattice of size $n \times m$ with a total of $M = n \times m$ cells. Let $X_i$ denote a random variable associated with cell $i$ such that $X_i \in \{1,2,\ldots, K\}$. Without loss of generality, in the rest of this work we will refer to these values or categories as colors. If $\mathbf{X}$ represents the entire random field, the probability of observing a particular configuration $\mathbf{x}$ is
    \begin{equation}\label{eq:PottsModel}
        P(\mathbf{X}=\mathbf{x})=\frac{\exp\{\sum_{k=1}^{K-1}\alpha_k T_{k} (\mathbf{x})+ \beta S(\mathbf{x})\}}{\sum_{\mathbf{x} \in \Omega}\exp\{\sum_{k=1}^{K-1}\alpha_k T_{k} (\mathbf{x})+ \beta S(\mathbf{x})\}},
    \end{equation}
  where
    \begin{equation*}
        T_{k}(\mathbf{x})=\sum_{i}I(x_i=k) \quad S(\mathbf{x})=\sum_i \sum_{i \sim j} I(x_i=x_j),
    \end{equation*}
and $i \sim j$ means that cells $i$ and $j$ are neighbors.

 The parameters in Equation (\ref{eq:PottsModel}) are interpreted as follows: $\boldsymbol{\alpha}=(\alpha_1,\ldots,\alpha_k\ldots ,\alpha_{K-1})^t$ is known as the external field parameter. We have implicitly set $\alpha_K=0$, so that color K is the reference color. Each element $\alpha_k$ controls the proportion of cells of color $k$ with respect to the proportion of cells of the color of reference $K$. Then, if $\alpha=\mathbf{0}$, all the colors are equally likely to occur in each cell. The parameter $\beta$ controls the spatial dependence between sites. When $\beta=0$, the random variables associated with the sites are independent, whereas large positive values of $\beta$ encourage neighboring sites to take the same color. Negative values of $\beta$ indicate repulsion between sites. In this work, we will focus exclusively on positive values of $\beta$. Finally, the denominator in Equation (\ref{eq:PottsModel}) corresponds to a normalizing function of the parameters $\alpha$ and $\beta$ that becomes intractable as the size of the lattice grows.  
 In Equation (\ref{eq:PottsModel}), there are also $K$ sufficient statistics. $T_k$ represents the number of pixels of color $k$ in the lattice, while $S$ measures the number of concordant neighbor pairs over discordant pairs. Examples of realizations from the Potts model for different the values of the parameters can be found in the supplementary materials.

\subsection{Phase transition and ground states}\label{sec-PhaseTransition}

This section investigates the behavior of the Potts model through simulations as its parameters vary. This analysis is particularly important because, for certain values of $\beta$, it is well-known that the Potts exhibits behaviors that become challenges in statistical modeling. In high-correlation settings, when the external field is absent (i.e., when $\boldsymbol{\alpha} = (0, \ldots, 0)^t$), the Ising model and the Potts model exhibit a characteristic known as phase transition \citep{Potts1952}, \citep{Wu1982}. In the physics context, a large value of $\beta$ translates into a low temperature. When the temperature is lower than a critical threshold, the system goes through a phase transition and it degenerates to a ground state \citep{Georgii2011}. In practical terms, this means that when the value of $\beta$ is close to or larger than the phase transition value $\log(1+\sqrt{K})$, the model places all its probability mass on extreme arrangements where most of the cells of the lattice are the same color. In the particular case of the Ising model, the model will randomly generate an arrangement with one of the two colors as the predominant color. A natural consequence of this behavior is multimodality and high variance in the distribution of $T(\mathbf{x})$.

\begin{figure}
    \centering
    \includegraphics[width=9.5cm, height=5cm]{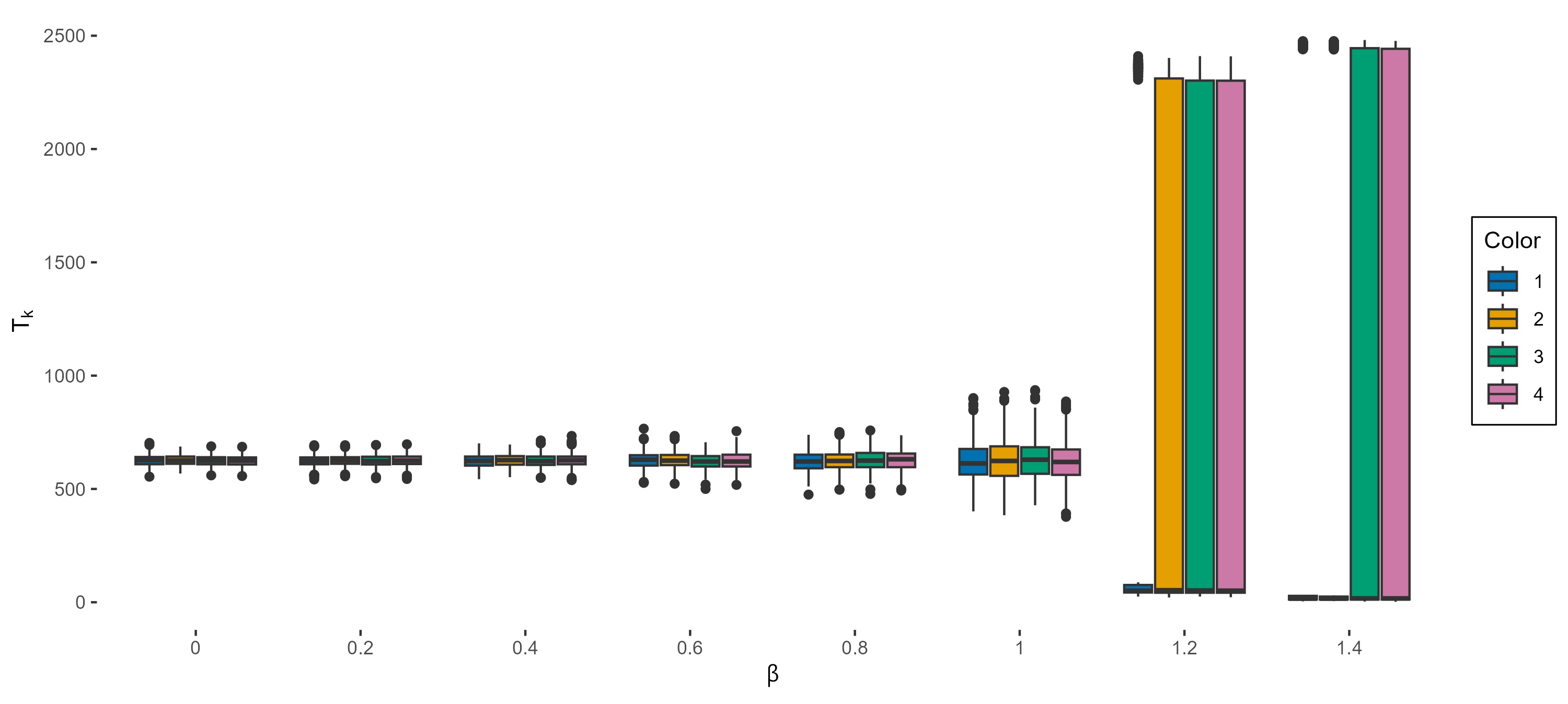}
    \includegraphics[width=5cm, height=5cm]{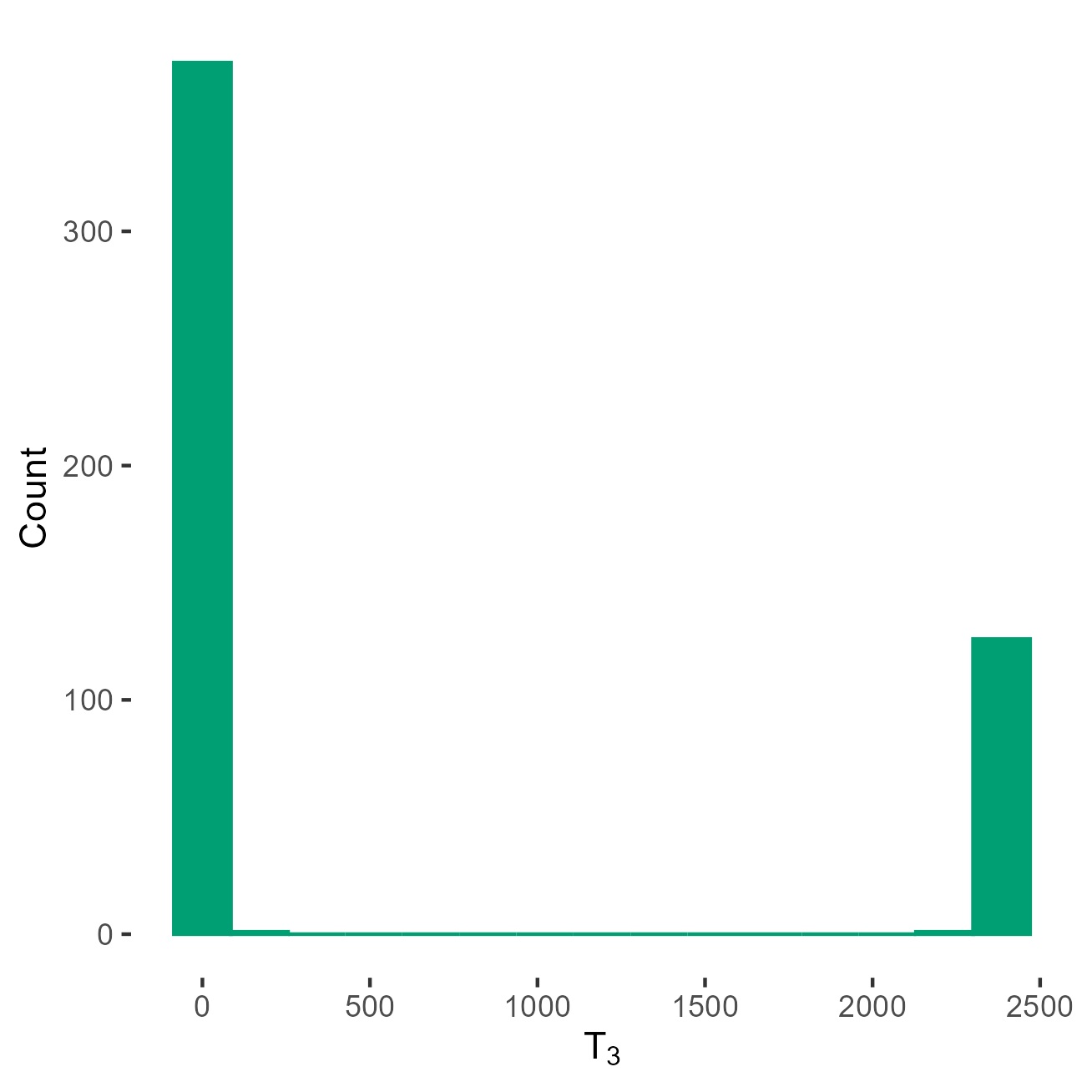}
    \caption{Boxplot of $T_{k}(\mathbf{x})$ based on a sample from a Potts model with different values of $\beta$ (left) and distribution of $T_{3}(\mathbf{x})$ from the Potts model when it has reached the phase transition value (right) when $\boldsymbol{\alpha}=(0,0,0)^t$.}
    \label{fig:BoxplotPottsNoExternal}
\end{figure}

The Potts model exhibits a similar behavior. Figure \ref{fig:BoxplotPottsNoExternal} (left) presents the results of computing the value of $T_k(\mathbf{x})$, $k = 1, \ldots, 4$, based on 500 arrangements from the Potts model parametrized by different values of $\beta$ and no external field. For $\beta \leq 1$, we observe that all four colors are present in roughly equal proportions across the simulated arrangements. However, for $\beta=1.2$ and $\beta=1.4$ we see an increase in the variance of the statistics $T_{k}(\mathbf{x})$. The shape of the boxplots in these cases indicates that the model is generating extreme arrangements where the majority of the cells have the same color. Figure \ref{fig:BoxplotPottsNoExternal} (right) zooms into the empirical distribution of $T_{3}(\mathbf{x})$ in this particular scenario. Since $K=4$, we observe that approximately 25\% of the arrangements are predominantly of color 3, while the remaining 75\% of the sample have very few cells of this color.  We conclude that, under high spatial correlation, the model tends to randomly choose one color to dominate in the resulting arrangement.

When $\boldsymbol{\alpha} \neq (0, \ldots, 0)^t$, we say that the Potts model has an asymmetrical external field \citep{Ahn2024}. In this case, and in the presence of high spatial correlation, if two or more elements of the vector $\boldsymbol{\alpha}$ are equal, it randomly chooses one of them to be the predominant color in the generated configuration. If one particular color $j$ is preferred, the arrangements become predominantly of color $j$. To our knowledge, this case has not been studied in the statistical modeling literature. 

\begin{figure}
    \centering
    \includegraphics[height=5cm]{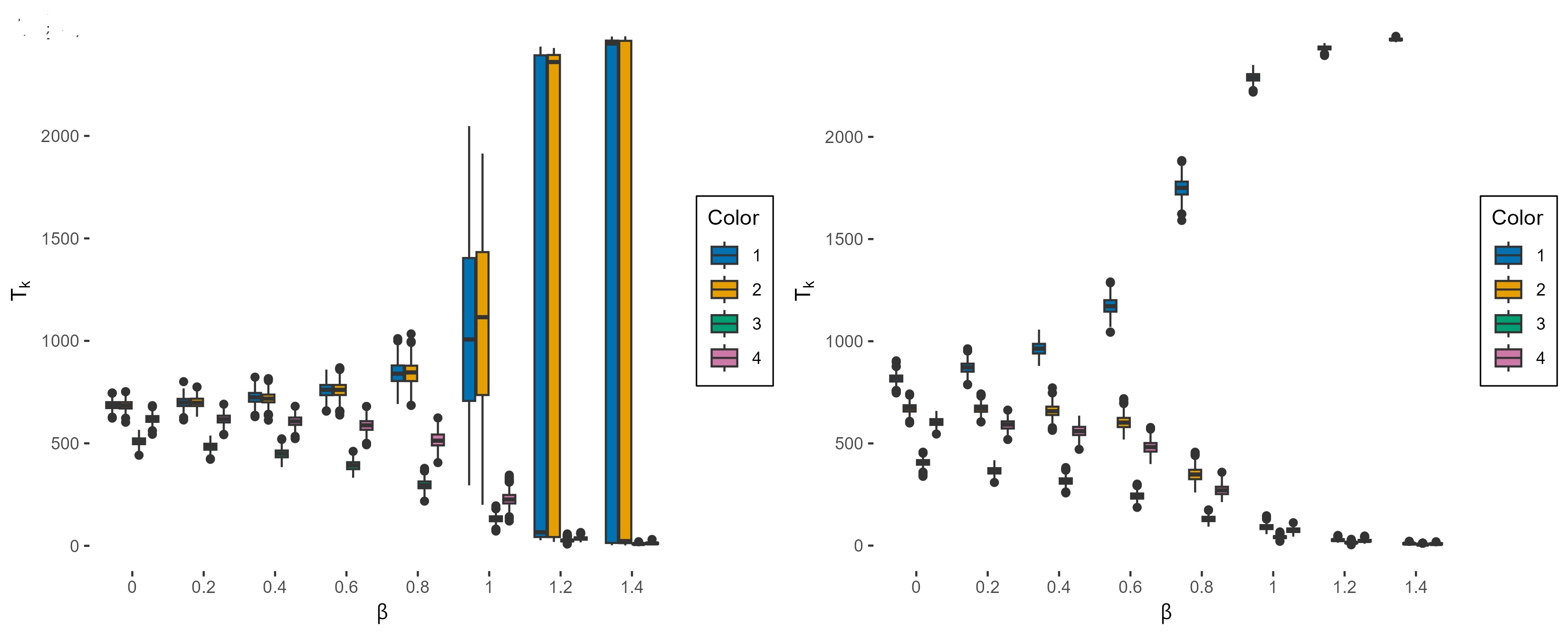}
    \caption{Boxplot of $T_{k}(\mathbf{x})$ based on a sample from a Potts model with different values of $\beta$ when $\boldsymbol{\alpha}=(0.1, 0.1, -0.2)^t$ (left) and $\boldsymbol{\alpha}=(0.3, 0.1, -0.4)^t$ (right).}
    \label{fig:BoxplotPottsExternal}
\end{figure}

The behavior of the Potts model under these particular settings is illustrated in Figure \ref{fig:BoxplotPottsExternal}. The plot on the left presents the empirical distribution of the statistics $T_{k}(\mathbf{x})$ for different values of $\beta$ when $\boldsymbol{\alpha}=(0.1, 0.1, -0.2)^t$. As colors 1 and 2 are equally preferred in this model, the distributions of $T_{1}(\mathbf{x})$ and $T_{2}(\mathbf{x})$ present multimodality for high values of $\beta$. In the meantime, colors 3 and 4 are present in an extremely small proportion when the spatial correlation is high. The plot on the right presents the empirical distribution of the statistics $T_{k}(\mathbf{x})$ for different values of $\beta$ when $\boldsymbol{\alpha}=(0.3, 0.1, -0.4)^t$. Under this parametrization, and if the value of $\beta$ is high, color 1 is preferred and all arrangements will have a majority of cells of color 1, while they all have a very small number of cells of colors 2,3, and 4.

\subsection{Undesirable characteristics of the Potts model}\label{sec-UndesirablePotts}

\begin{figure}
    \centering
    \includegraphics[width=5cm, height=5cm]{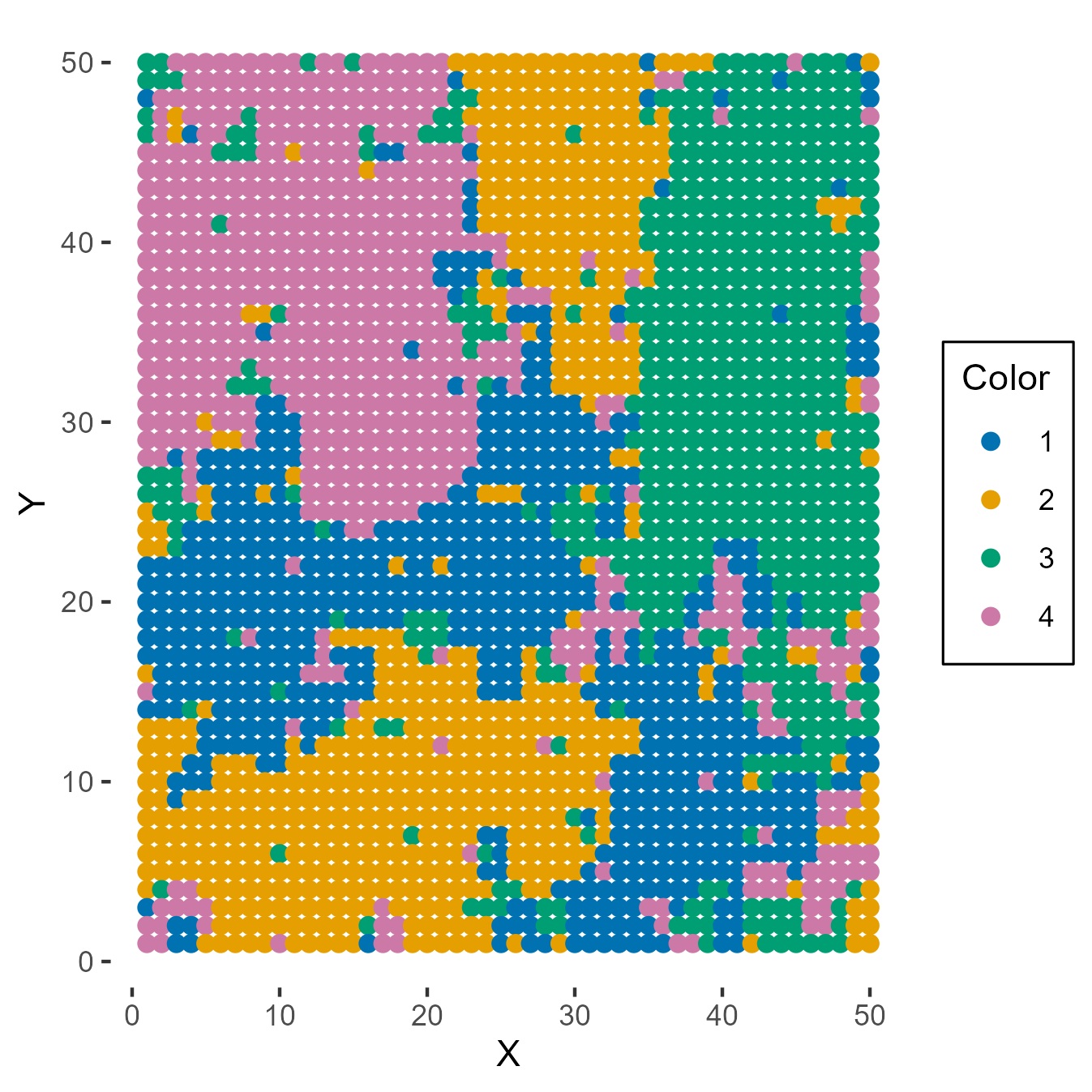}
    \includegraphics[width=5cm, height=5cm]{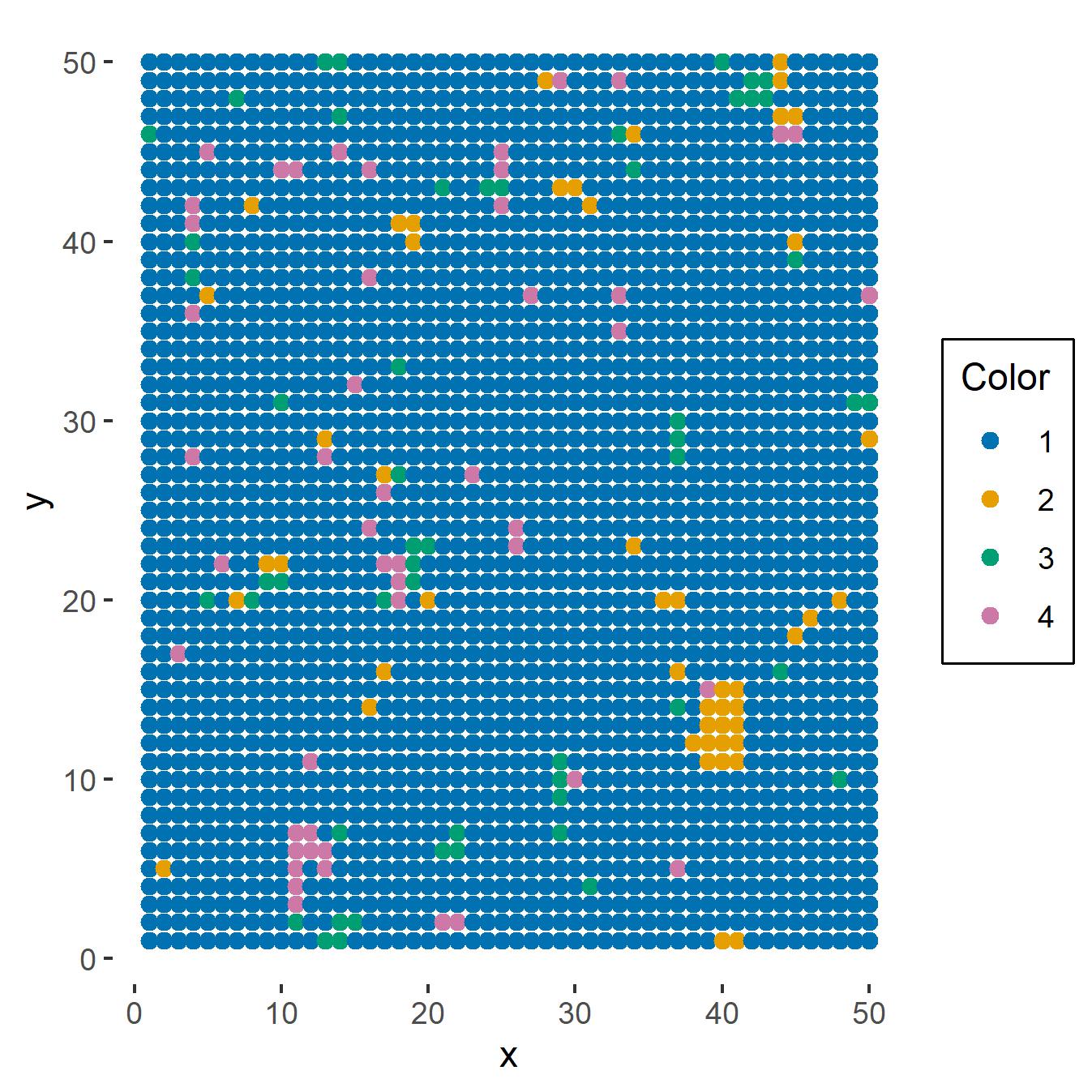}
    \caption{Unlikely configuration under the Potts model (left) and likely configuration under the Potts model (right)}
    \label{fig:NoExternalFieldUnlikely}
\end{figure}

Reaching ground states in the presence of high spatial correlation creates challenges when modeling real datasets of multi-category lattice data using the Potts model, as there is a wide range of datasets with high spatial correlation that also have a significant presence of different colors in their cells. However, when the Potts model reaches its ground state arrangement like these are unlikely. Figure \ref{fig:NoExternalFieldUnlikely} (left) presents an example of such arrangements, while Figure \ref{fig:NoExternalFieldUnlikely} (right) presents an example of a typical realization of the Potts model under settings of high spatial correlation.

Using the Potts model to fit arrangements like the one in Figure \ref{fig:NoExternalFieldUnlikely} would result in a lack of fit, and fitted models would be unable to simulate new arrangements that reproduce the properties of the original data.

\subsection{Maximum likelihood inference}\label{sec-MLEPotts}

Let $\boldsymbol{\theta}=(\alpha_1,\ldots,\alpha_{k-1}, \beta)$ and $\mathbf{G}(\mathbf{x}^{obs})=\left(T_{1}(\mathbf{x}^{obs}),\ldots, T_{k-1}(\mathbf{x}^{obs}), S(\mathbf{x}^{obs})\right)$. Then, the log-likelihood for the Potts model can be written as
\begin{equation}\label{eq:originalloglik}
    l(\boldsymbol{\theta})=\log\left(L(\mathbf{x}^{obs}, \boldsymbol{\theta})\right)=\boldsymbol{\theta}^t \mathbf{G}(\mathbf{x}^{obs})-\log\left(c(\boldsymbol{\theta})\right).
\end{equation}
Given that $c(\boldsymbol{\theta})$ is an intractable normalizing function of the parameters, the log-likelihood function in (\ref{eq:originalloglik}) is difficult to maximize. However, one approach to avoid this issue is approximating the log-likelihood function using simulations obtained with Markov Chain Monte Carlo (MCMC). This method is known as Monte Carlo Maximum Likelihood (MCMCMLE) \citep{Geyer1991}. This approach fixes an arbitrary value of the parameter $\boldsymbol{\theta}_0$ to approximate the log-likelihood ratio $\log\left(\frac{L(\mathbf{x}^{obs}, \boldsymbol{\theta})}{L(\mathbf{x}^{obs}, \boldsymbol{\theta}_0)}\right)$. \cite{Geyer1991} showed that the log-likelihood ratio can be written as 
\begin{equation} \label{eq:expectedval}
    l(\boldsymbol{\theta})-l(\boldsymbol{\theta}_0)=(\boldsymbol{\theta}-\boldsymbol{\theta}_0)^t\mathbf{G}(\mathbf{x}^{obs})-\log \left(E_{\boldsymbol{\theta}_0}\left[\exp \{(\boldsymbol{\theta}-\boldsymbol{\theta}_0)^t \mathbf{G}(\mathbf{Y})\}\right]\right),
\end{equation}
and the expected value in Equation (\ref{eq:expectedval}) can be approximated by sampling from the distribution parametrized by $\boldsymbol{\theta}_0$. Section \ref{sec-PottsSimulation} mentions methods for simulating realizations from the Potts model, and more details on how to approximate maximum likelihood estimates using MCMCMLE for the Potts model can be found in the supplementary materials.

While $\boldsymbol{\theta}_0$ can be chosen arbitrarily in theory, it is crucial to find a value of $\boldsymbol{\theta}_0$ close enough to $\hat{\boldsymbol{\theta}}$ to reasonably approximate the log-likelihood ratio and the real value of the maximum likelihood estimates. \cite{Hummel2012} proposed the partial stepping algorithm, which moves $\boldsymbol{\theta}_0$ closer to $\hat{\boldsymbol{\theta}}$ using exponential-family theory. Details of this algorithm can be found in their paper and the supplementary materials.

\subsubsection{Simulating from the Potts model} \label{sec-PottsSimulation}

One classical approach for simulating realizations of the Potts model is using a Gibbs sampler. However, when $\beta$ exceeds the phase transition value, the algorithm tends to stay in one of the modes or ground states, leading to inaccurate sample averages. \cite{Geyer1991} suggests using a symmetric swap of the values at each site inside of the Gibbs sampler to mitigate this issue in the classic Ising model. We generalized this approach to the classic Potts model. The complete Gibbs algorithm is described in the supplementary materials. On the other hand, the Swendsen-Wang algorithm \citep{Swendsen1987} is a popular alternative that offers efficiency and better mixing properties when simulating samples from the Potts model. Details on how to build this algorithm can be found in \cite{GeyerLectureNotes}. In this work, we used the Swendsen-Wang algorithm implemented in the \textit{potts} R package \citep{pottslibrary}.

\subsubsection{Identifying lack of fit of the Potts model}\label{sec-LackofFit}

In Section \ref{sec-PhaseTransition} we describe the ground states of the Potts model and how these affect its fit to datasets with high levels of spatial correlation present, such as the one in Figure \ref{fig:NoExternalFieldUnlikely}. In this section, we provide a heuristic using the partial stepping algorithm of \cite{Hummel2012} to identify these situations when the Potts model is a poor fit to a data set, thereby necessitating the use of the tapered Potts model we propose in the next section. In particular, the partial stepping algorithm attempts to get close to a value of $\boldsymbol{\theta}_0$ that could generate a sample mean equal to $\mathbf{G}(\mathbf{x}^{obs})$. However, if the model cannot reproduce a sample mean close to the observed sufficient statistics because it has reached the phase transition value, the partial stepping algorithm does not converge and we will not find an approximation to the maximum likelihood estimate. In those cases, we will need an alternative to the Potts model.

\section{The Tapered Potts model}\label{sec-TapPotts}

To address the lack-of-fit issues described in the previous sections, we propose augmenting the Potts model with an additional term. This term mitigates the tendency of the Potts model to degenerate into ground states and instead encourages the generation of arrangements whose observed sufficient statistics are close to those observed in the original dataset. In this way, we will generate new configurations that maintain a composition similar to that of the original arrangement, which is our main goal. Our methodology is inspired by the work of \cite{Handcock2017} in the context of exponential random graph models (ERGM). 

In order to improve the fit of the model in the presence of high spatial correlation, we are interested in constraining the distributions of the statistics $T_k(\mathbf{x})$ around $T_k(\mathbf{x}^{obs})$ . Therefore, we consider the next optimization problem:
\begin{align*}
\text{maximize} & \sum_{x}q(x)\log\left(q(x)\right), \\
  \text{subject to} & \sum_{x} q(x)=1,  \\
  & E\left(T_k(\mathbf{x})\right)=T_k(\mathbf{x}^{obs}),\\
  & E\left(\left(T_k(\mathbf{x})-T_k(\mathbf{x}^{obs})\right)^2\right) \leq \kappa_k,  \quad k=1,\ldots, K-1.
\end{align*}

Using Karush-Kuhn-Tucker multipliers, \cite{Handcock2017} showed that the maximum entropy distribution subject to variability constraints is
\begin{equation} \label{eq:tap_potts}
    q(x|\alpha, \beta, \tau)=\frac{1}{c(\alpha, \beta, \tau)}\exp\biggl\{\sum_{k=1}^{K-1}\alpha_k T_{k} (\mathbf{x})+ \beta S(\mathbf{x})-\sum_{k=1}^{K-1}\tau_k\left(T_{k}(\mathbf{x})-T_k(\mathbf{x}^{obs})\right)^2\biggl\},
\end{equation}
where $c(\alpha, \beta, \tau)$ is an intractable normalizing function. Note that the distribution in Equation \ref{eq:tap_potts} belongs to the exponential family.

The new term $\sum_{k=1}^{K-1}\tau_k\left(T_{k}(\mathbf{x})-T_k(\mathbf{x}^{obs})\right)^2$ will be called the \textit{tapering term} for the rest of this work. The new parameter $\boldsymbol{\tau}=(\tau_1, \ldots, \tau_{K-1})$ controls the level of tapering for each sufficient statistic $T_{k}$. The algorithm for simulating from the tapered Potts model in Equation (\ref{eq:tap_potts}) results from implementing a few changes in the classic Gibbs sampler for the Potts model. Details can be found in the supplemental materials.

\subsection{Maximum likelihood inference for the tapered Potts model}\label{sec-MLTapPotts}

By the conjugate duality property that relates maximum entropy and maximum likelihood in the exponential family \citep{Wainwright2008} we have that $\hat{\theta}_{ML}$ is the value such 
    \begin{equation}
        E_{\boldsymbol{\widehat\theta}_{ML}}\left[\mathbf{T}(\mathbf{x})\right]=\mathbf{T}(\mathbf{x}^{obs}).
    \end{equation}
Using this fact, we can find $\hat{\theta}_{ML}$ applying the MCMCMLE approach described in Section \ref{sec-MLEPotts}. We will only need to generate samples from a Tapered Potts model instead of the classical Potts model. Because there is no information on the parameter $\boldsymbol{\tau}$ in the likelihood when $\theta=\hat{\theta}_{ML}$, its value must be fixed before generating samples from the Tapered model. Adequate values of $\boldsymbol{\tau}$ can be found using a numerical approach described in Section \ref{sec-ChooseTau}.

\subsection{Effects of tapering in the Potts model}

\begin{figure}
    \centering
    \includegraphics[width=5cm, height=5cm]{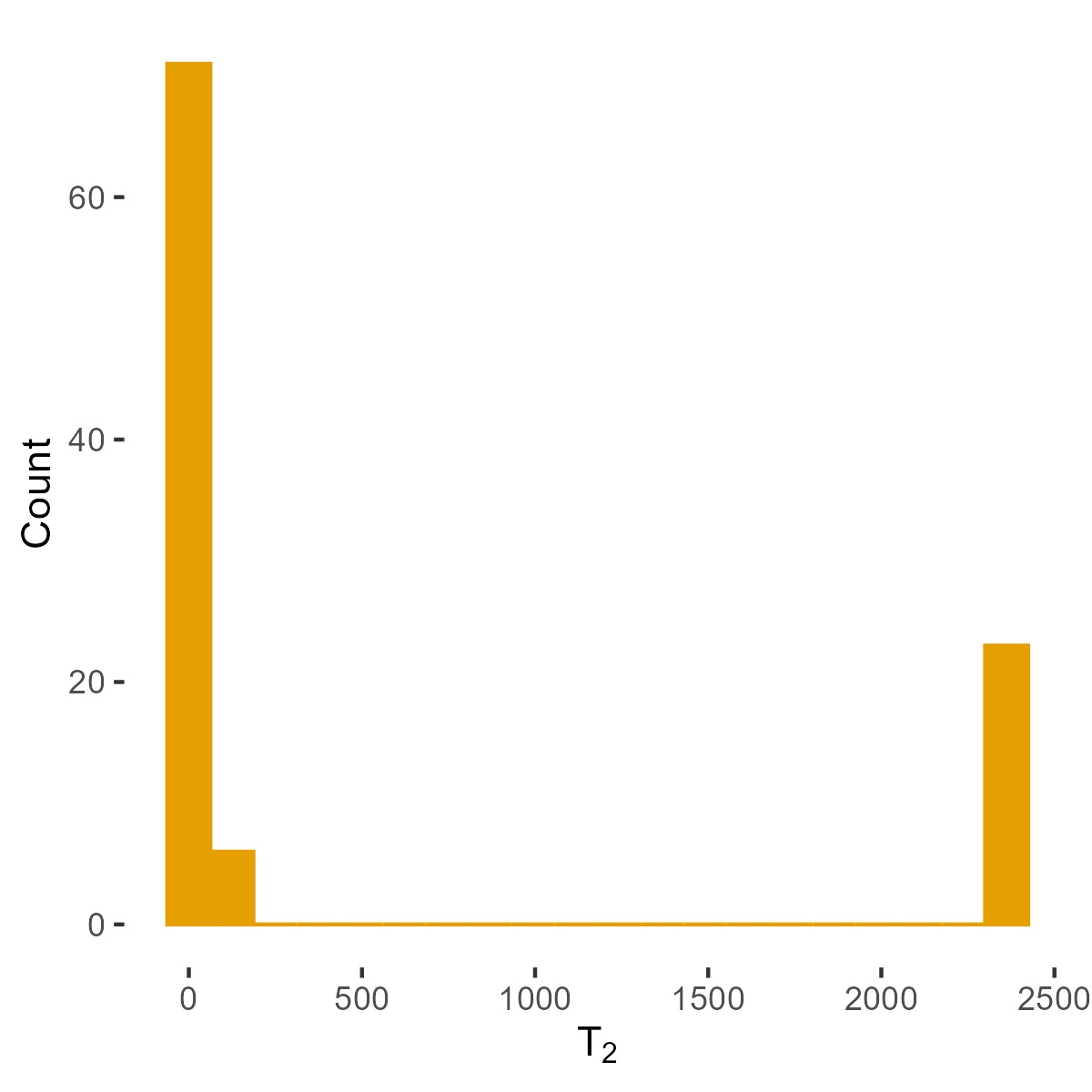}
    \includegraphics[width=5cm, height=5cm]{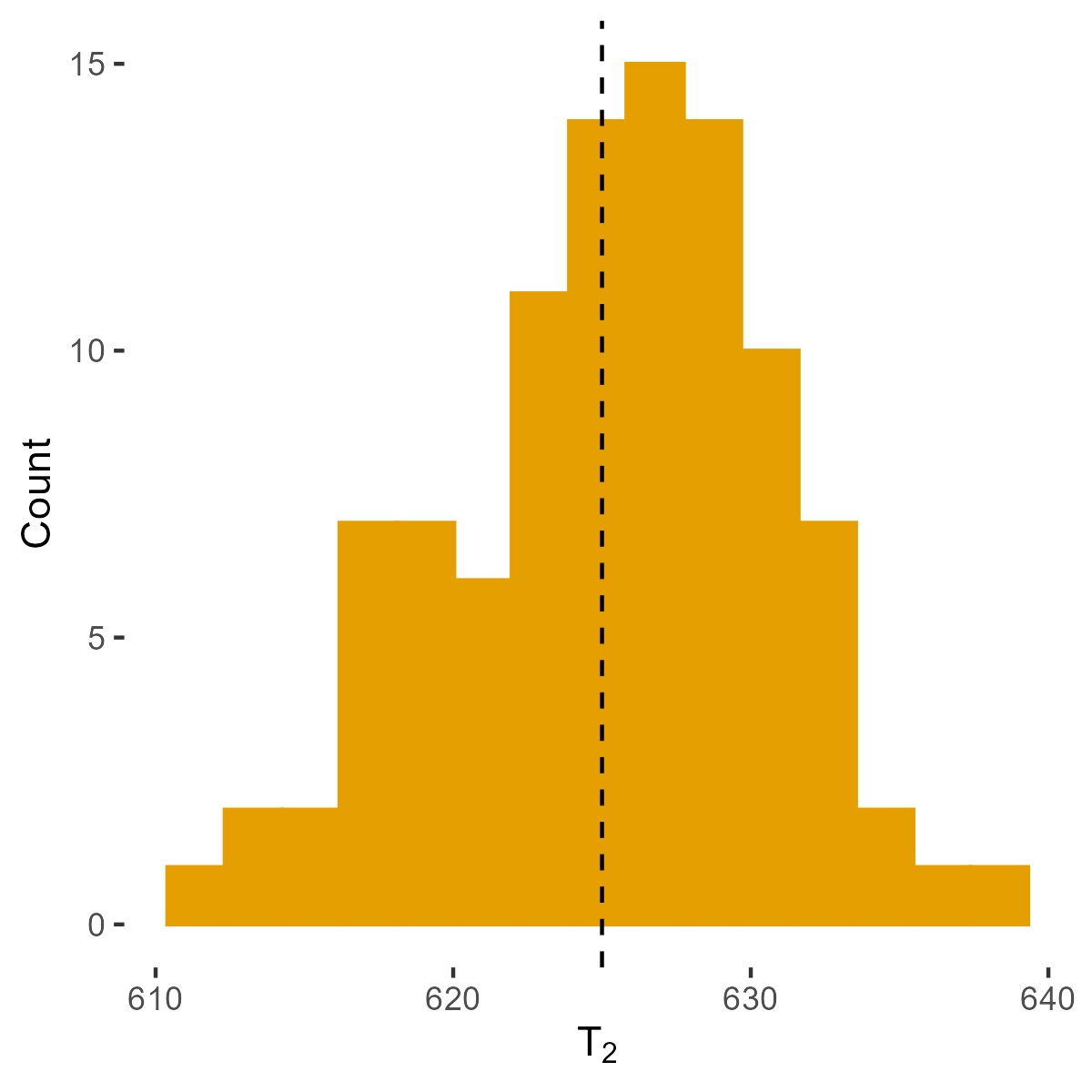}
    \caption{Distribution $T_{k}(\mathbf{x})$ when $\beta$ exceeds the phase transition value (left) and effect of tapering in the distribution of $T_{k}(\mathbf{x})$ (right) when $\boldsymbol{\alpha}=0$.}
    \label{fig:effectTap_noexternal}
\end{figure}

This section portrays the effect of adding a tapering term to the Potts model and how it can help manage the challenges described in Section \ref{sec-LackofFit}. We will consider two cases: $\boldsymbol{\alpha}=0$ and $\boldsymbol{\alpha} \neq 0$. In the particular case of $\boldsymbol{\alpha}=0$, Figure \ref{fig:effectTap_noexternal} (left) shows the distribution of a particular statistic $T_{k}(\mathbf{x})$, which becomes bimodal when $\beta$ exceeds the phase transition value. Figure \ref{fig:effectTap_noexternal} (right) demonstrates the effect of including the tapering term in the Potts model with $\tau_k$ large enough: the distribution of $T_{k}(\mathbf{x})$ becomes unimodal and is centered around $T_{k}(\mathbf{x}^{obs})$. However, it is not recommended to choose higher values of $\tau_k$ than necessary, as they could overconstrain the spread of the statistic. 

On the other hand, if $\boldsymbol{\alpha} \neq 0$ and the Potts model reaches its ground states, when $j$ is the preferred color according to $\boldsymbol{\alpha}$ (that is $\alpha_j=\max_{1,\ldots,K-1}\{\alpha_k\}$) the distribution of $T_{j}(\mathbf{x})$ concentrates its mass on full configurations where the majority of cells are of color $j$. If $\alpha_j \neq \max_{1,\ldots,K-1}\{\alpha_k\}$, then the distribution of $T_{j}(\mathbf{x})$ concentrates its mass on empty configurations where few cells are of color $j$. In those cases, for $\tau_j$ sufficiently large, the tapering term will move the center of the distribution around $T_{j}(\mathbf{x}^{obs})$. 

It is clear that choosing the value of $\tau_j$ is crucial to the application of the Tapered Potts model in real datasets. We provide practical guidance for this choice in subsequent sections.

\subsection{Choosing the tapering value}\label{sec-ChooseTau}

It is necessary to use a numerical approach to find an adequate value for it $\tau_j$. In particular, two different approaches are necessary to determine an appropriate level of tapering depending on the value of $\boldsymbol{\alpha}$. When $\boldsymbol{\alpha} = \boldsymbol{0}$, we are dealing with the multimodality of the distribution of $T_k$. This implies that the partial algorithm may converge because the sample averages of the sufficient statistics under the Potts model may resemble the observed values. However, the simulated realizations will still fail in capturing the characteristics of the original data, and we need to add a tapering term. When $\boldsymbol{\alpha} \neq \boldsymbol{0}$, the lack of convergence of the partial stepping algorithm will indicate the need for tapering.

Care must be taken when choosing this value because a small value of $\tau_j$ may not fix the lack of fit problems of the Potts model. However, we want to use the smallest possible value of $\tau_j$ since it reduces the spread of $T_{j}(\mathbf{x})$ as it grows. We present an approach to find an adequate value for $\tau_j$ when $\boldsymbol{\alpha}=0$ and when $\boldsymbol{\alpha} \neq 0$. We will only consider the case where $\tau_j=\tau$ for all $j=1,\ldots,K-1$.

     \begin{enumerate}
     \item Set a large value for $\tau$.
            \item Find the MLE estimate of $\boldsymbol{\alpha}$ and $\beta$.
            \item Generate a sample of arrangements $\mathbf{x}_1, \ldots, \mathbf{x}_l$ from the fitted model.
            \item If there is a similar proportion of cells of each color in the observed arrangement or in the sample of arrangements obtained from the fitted model, the real value of $\boldsymbol{\alpha}$ could be close to zero. In that case:
            \begin{enumerate}
                \item  Measure the bimodality of the distribution of $T_{j}(\mathbf{x})$ based on the sample obtained in step 2. To do so, use the bimodality coefficient developed by \cite{Handcock2023}:
            \begin{equation}\label{eq:bimodcoef}
                \lambda=\frac{\gamma_1^2+1}{\gamma_2}.
            \end{equation}
            In Equation (\ref{eq:bimodcoef}), $\gamma_1$ is the sample skewness of $T_{j}(\mathbf{x})$, and $\gamma_2$ is the sample kurtosis. \cite{Handcock2023} recommends considering a distribution to be unimodal when $\lambda \leq 5/9$, as this is the bimodality coefficient of the uniform distribution.
            \item If the distribution is unimodal, reduce the value of $\tau$.
            \item Continue until the distribution of $T_{j}(\mathbf{x})$ is no longer unimodal and take the smallest value of $\tau_j$ such that the distribution is unimodal.
            \end{enumerate}
           \item If there is evidence in the data that $\boldsymbol{\alpha} \neq 0$ and the partial stepping algorithm converges, reduce $\tau$. Stop when the partial stepping algorithm no longer converges and take the smallest value of $\tau$ such that the algorithm converges.
        \end{enumerate}

\section{Simulation study}\label{sec-simStudy}

We simulate from Gaussian processes to study the Potts and Tapered Potts models in scenarios with varying levels of spatial correlation on a $30 \times 30$ grid.

\begin{enumerate}
    \item We calculated the Euclidean distance between the centroids of the cells. The distance between the centroids of cell $i$ and cell $j$ will be denoted by $d_{ij}$.
    \item For $K=4$ classes, we simulated four different Gaussian processes $\mathbf{Z}_k$, $k=1,\ldots,4$ using the $\gamma-\text{exponential}$ kernel with length parameter $l$ to model spatial correlation. That is $\mathbf{Z}_k \sim GP(\boldsymbol{\mu}_i, \Sigma_l)$, $\Sigma_{(i,j)}=\exp\Bigl\{-\left(d_{ij}/l\right)^\gamma\Bigl\}$.
    \item We assign the value $k=\max_{i}\{Z_i, i=1,\ldots, 4\}$ to each cell.  
\end{enumerate}

We will consider six different simulation scenarios:
\begin{itemize}
    \item In scenarios 1, 2 and 3 there is no preference for a particular color as $\boldsymbol{\mu}_i=0\times \mathbf{1}_{900}$, $i=1\ldots,4$. There are three different levels of spatial correlation: moderate, strong spatial correlation with some noise, and strong spatial correlation with minimal noise.
    \item In scenarios 4, 5 and 6 there are different levels of preference for colors 1 and 2 by setting $\boldsymbol{\mu}_1=\mathbf{1}_{900}$, $\boldsymbol{\mu}_2=0.5 \times \mathbf{1}_{900}$, $\boldsymbol{\mu}_3=0\times \mathbf{1}_{900}$, and $\boldsymbol{\mu}_4=0\times \mathbf{1}_{900}$. We consider the same three levels of spatial correlation as in scenarios 1,2 and 3.
    
\end{itemize}

\begin{figure}
    \centering
    \includegraphics[width=5cm, height=3cm]{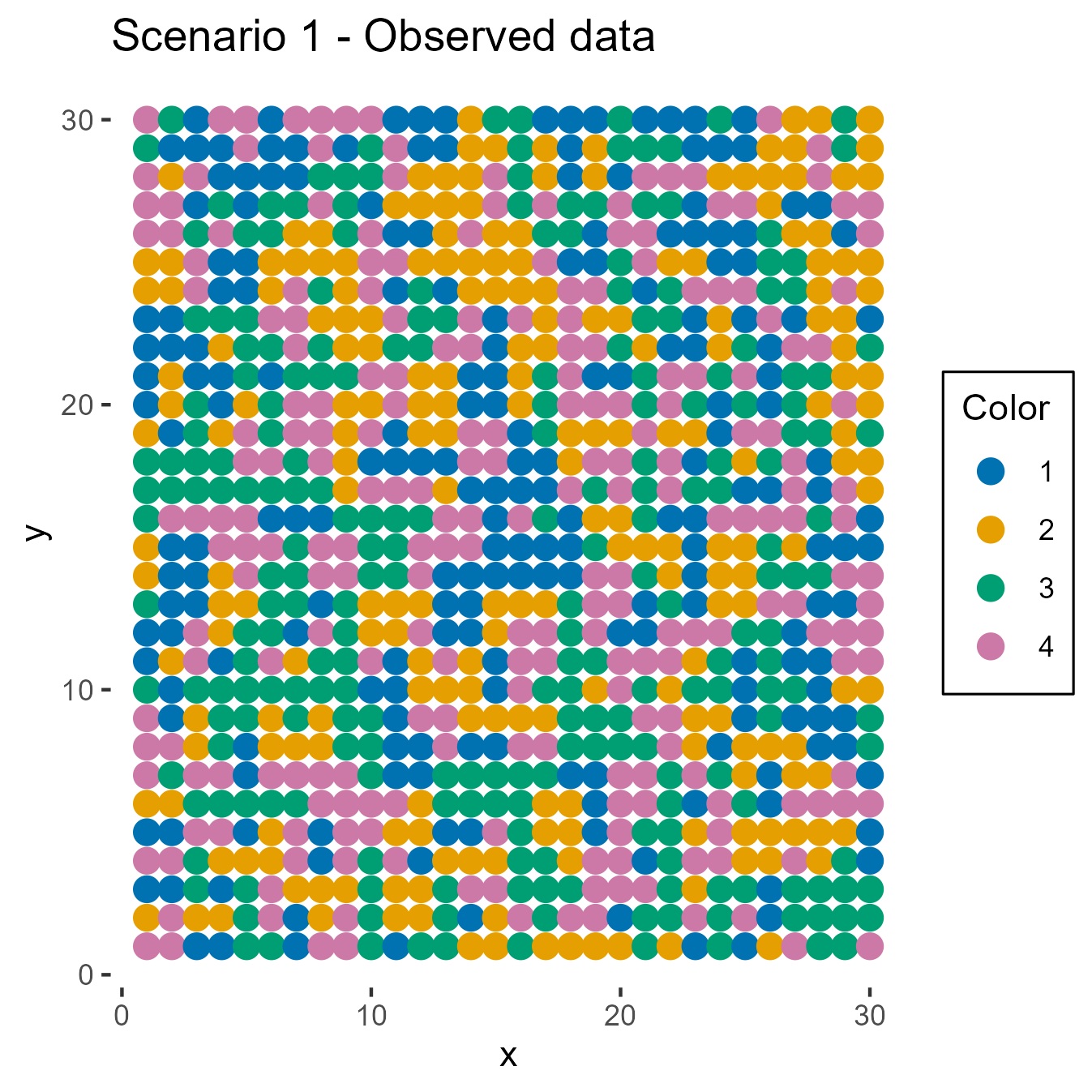}
    \includegraphics[width=5cm, height=3cm]{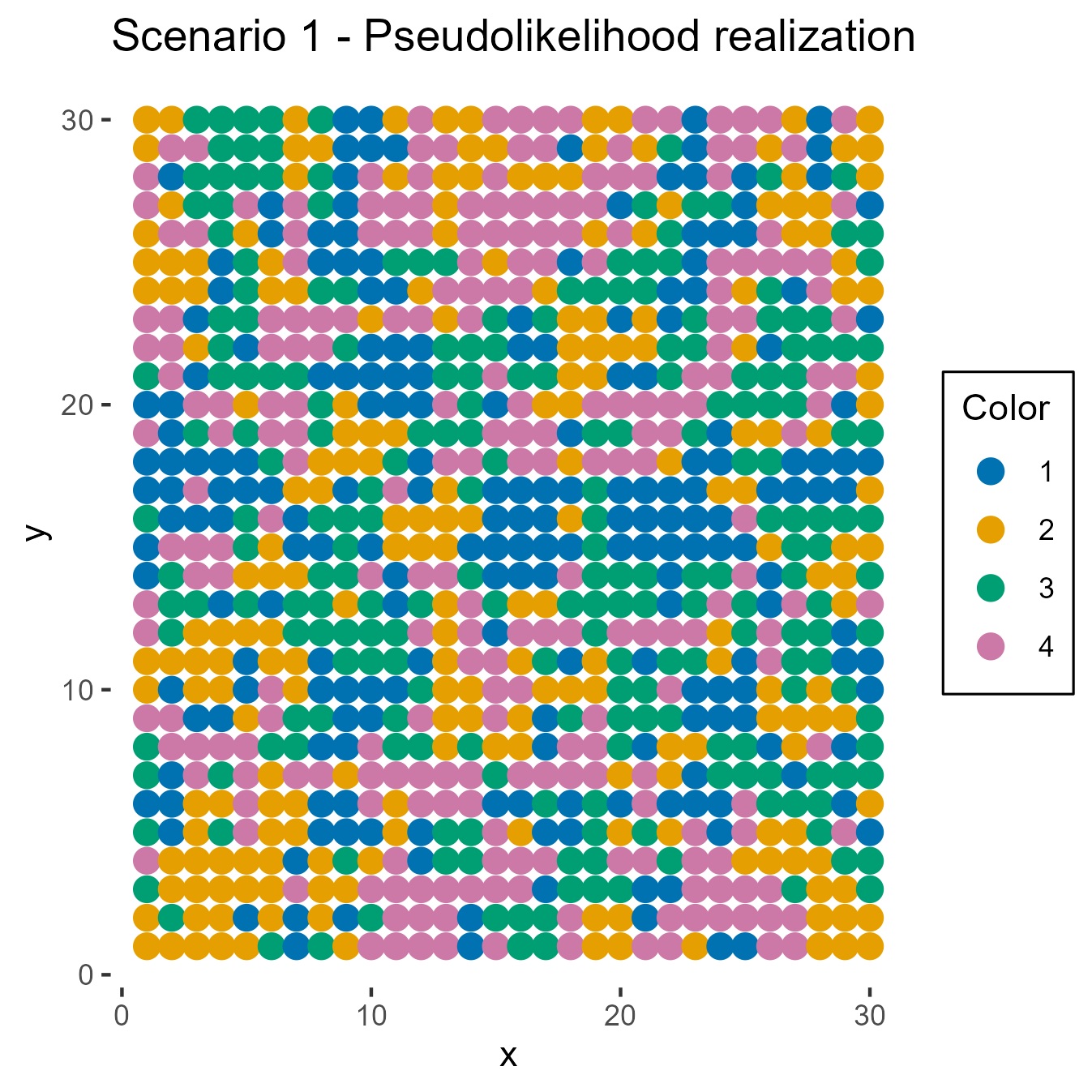} 
    \includegraphics[width=5cm, height=3cm]{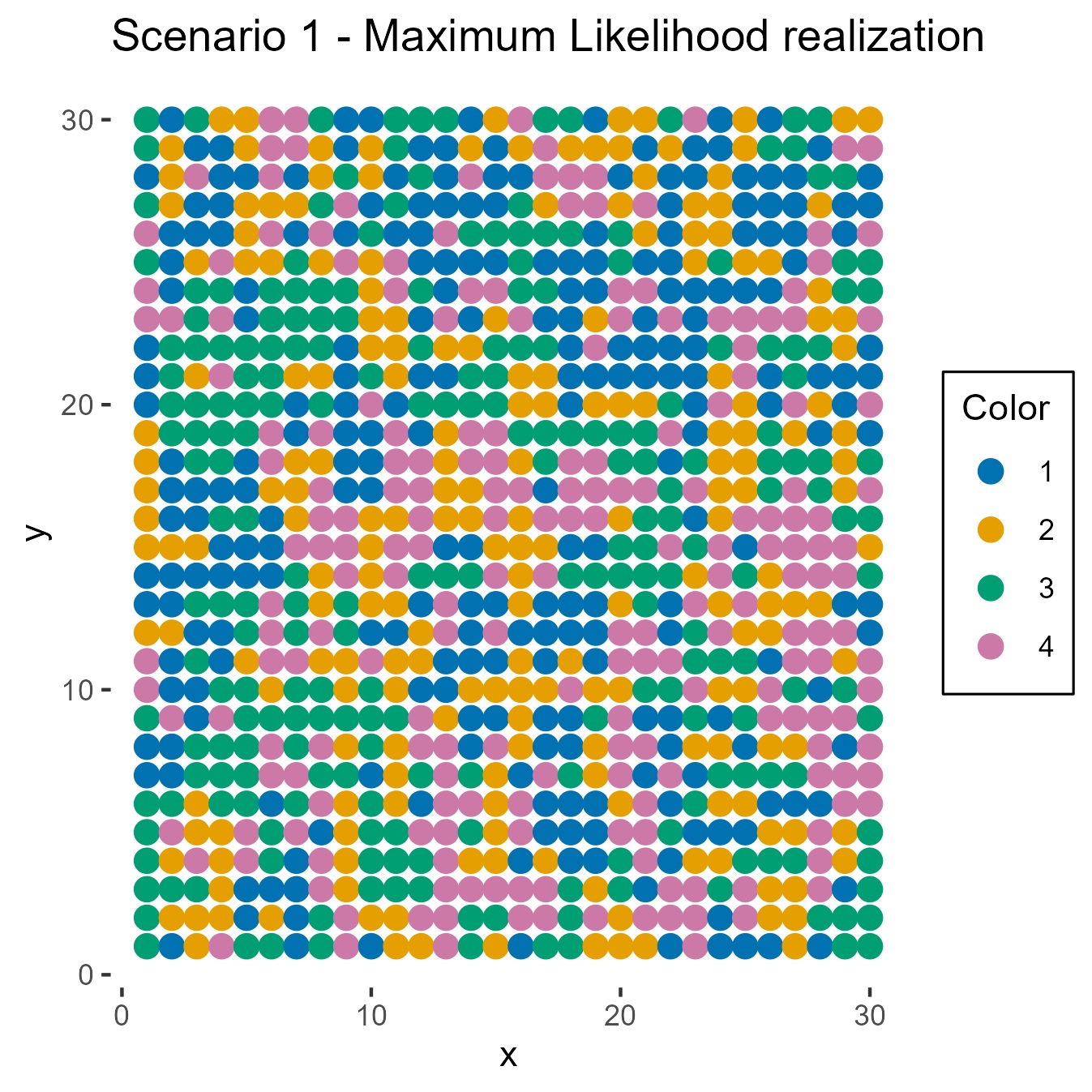} \\
    \includegraphics[width=5cm, height=3cm]{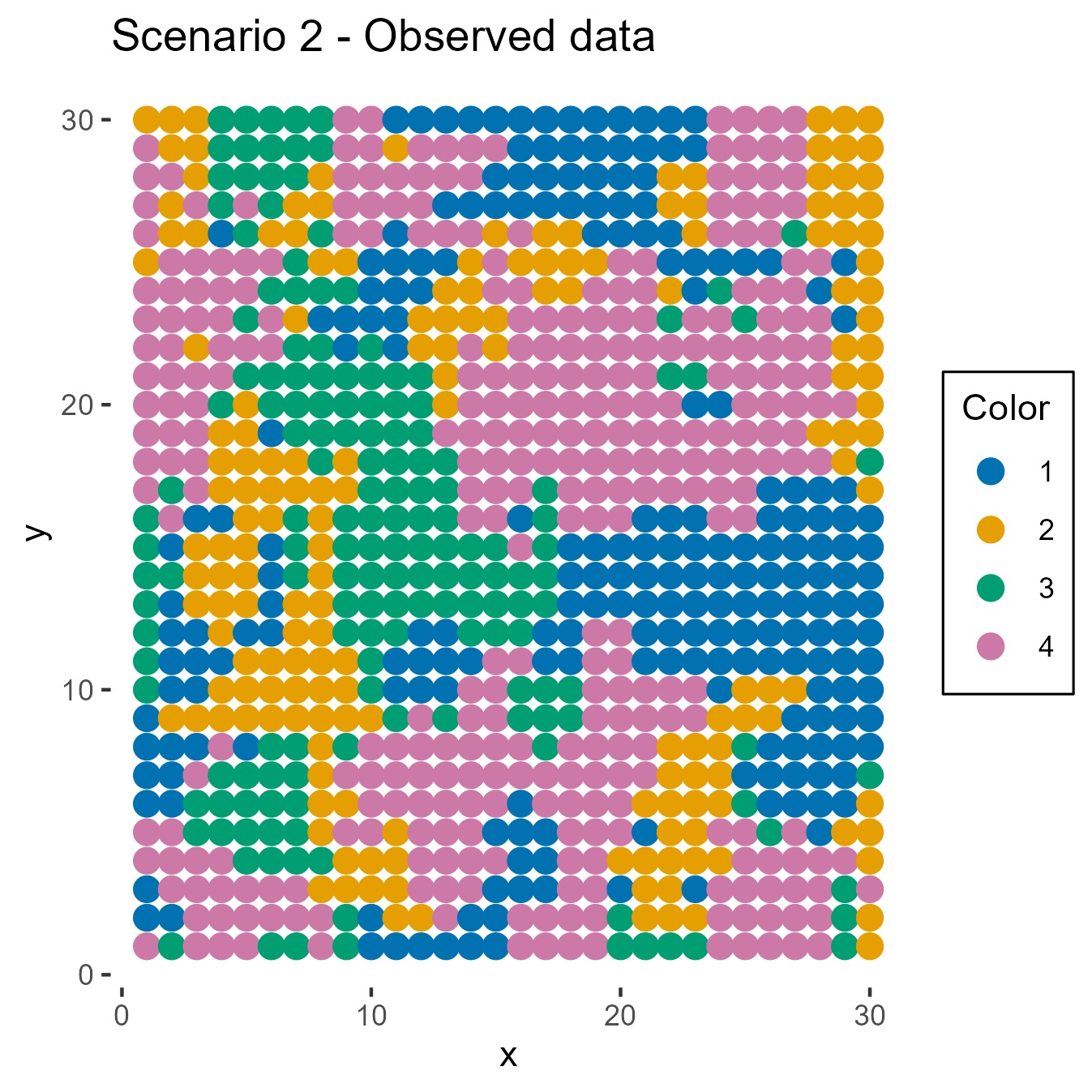}
    \includegraphics[width=5cm, height=3cm]{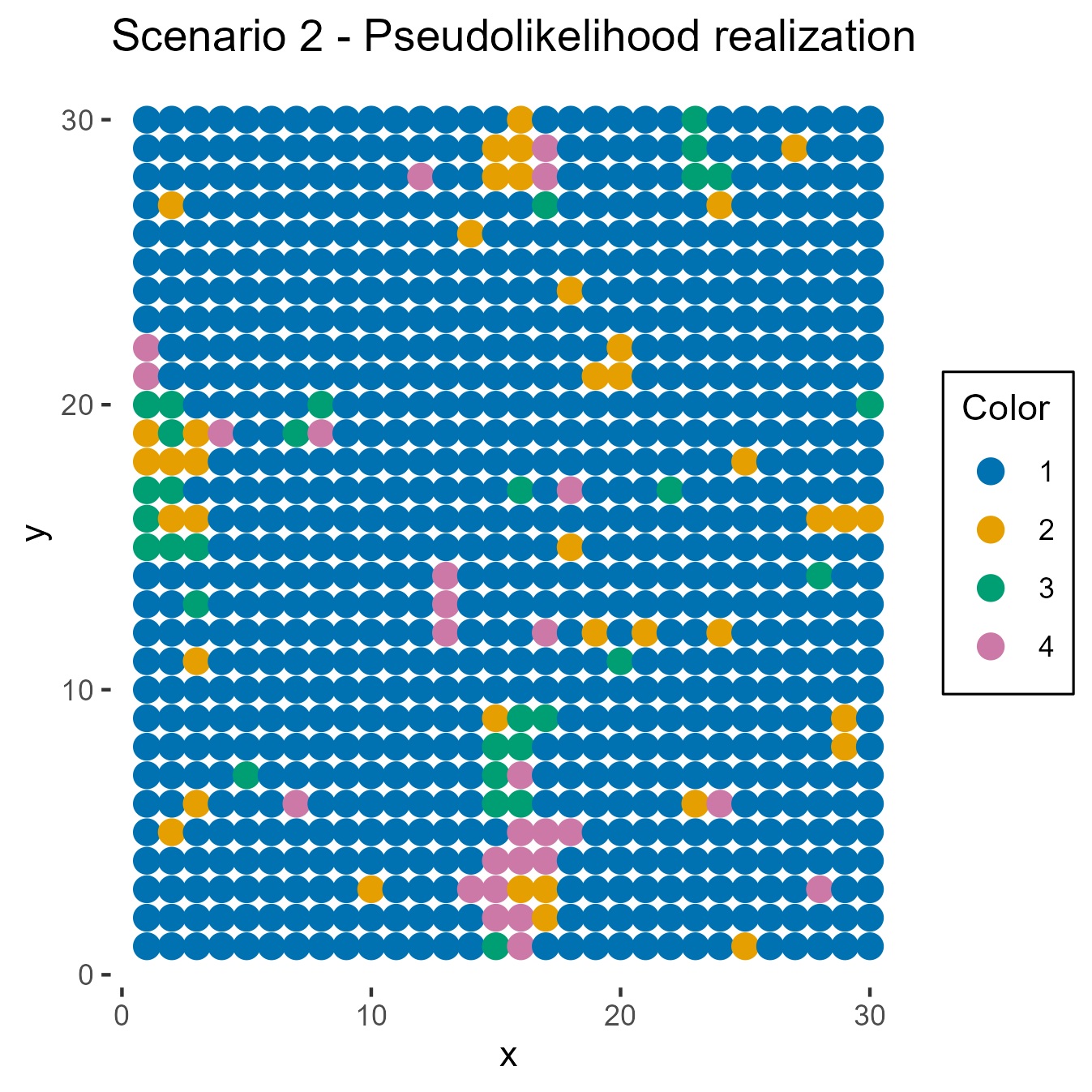}
    \includegraphics[width=5cm, height=3cm]{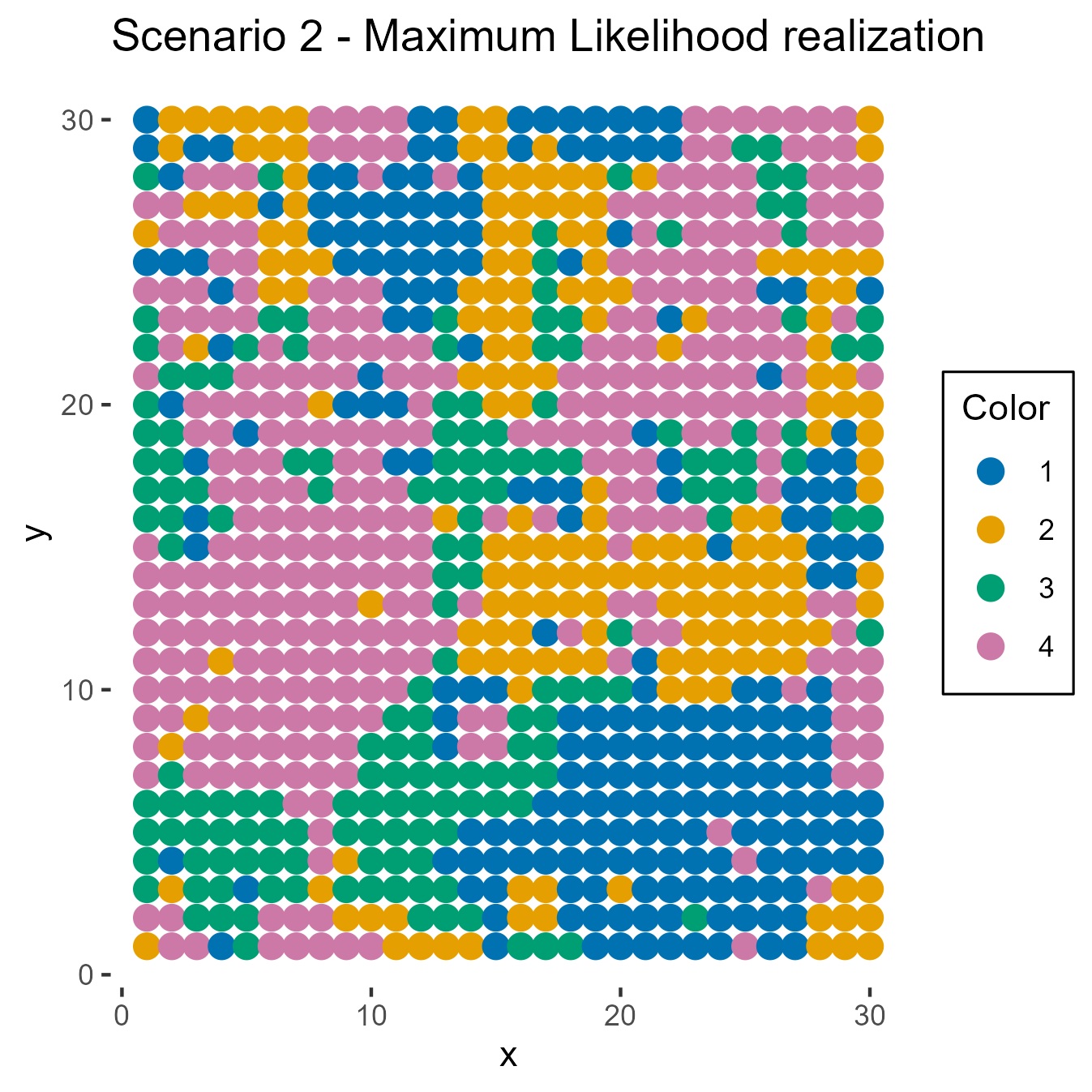}\\
    \includegraphics[width=5cm, height=3cm]{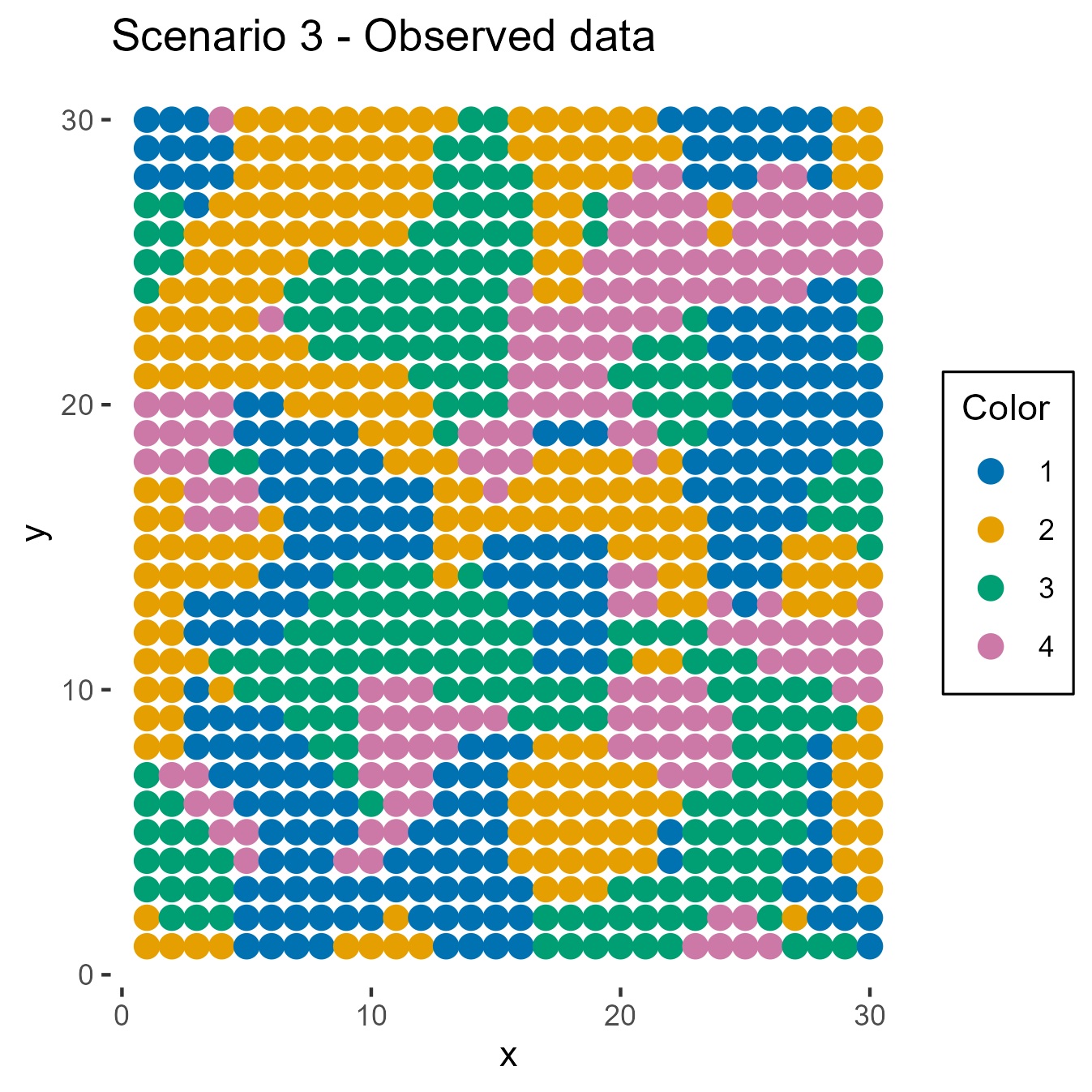}
    \includegraphics[width=5cm, height=3cm]{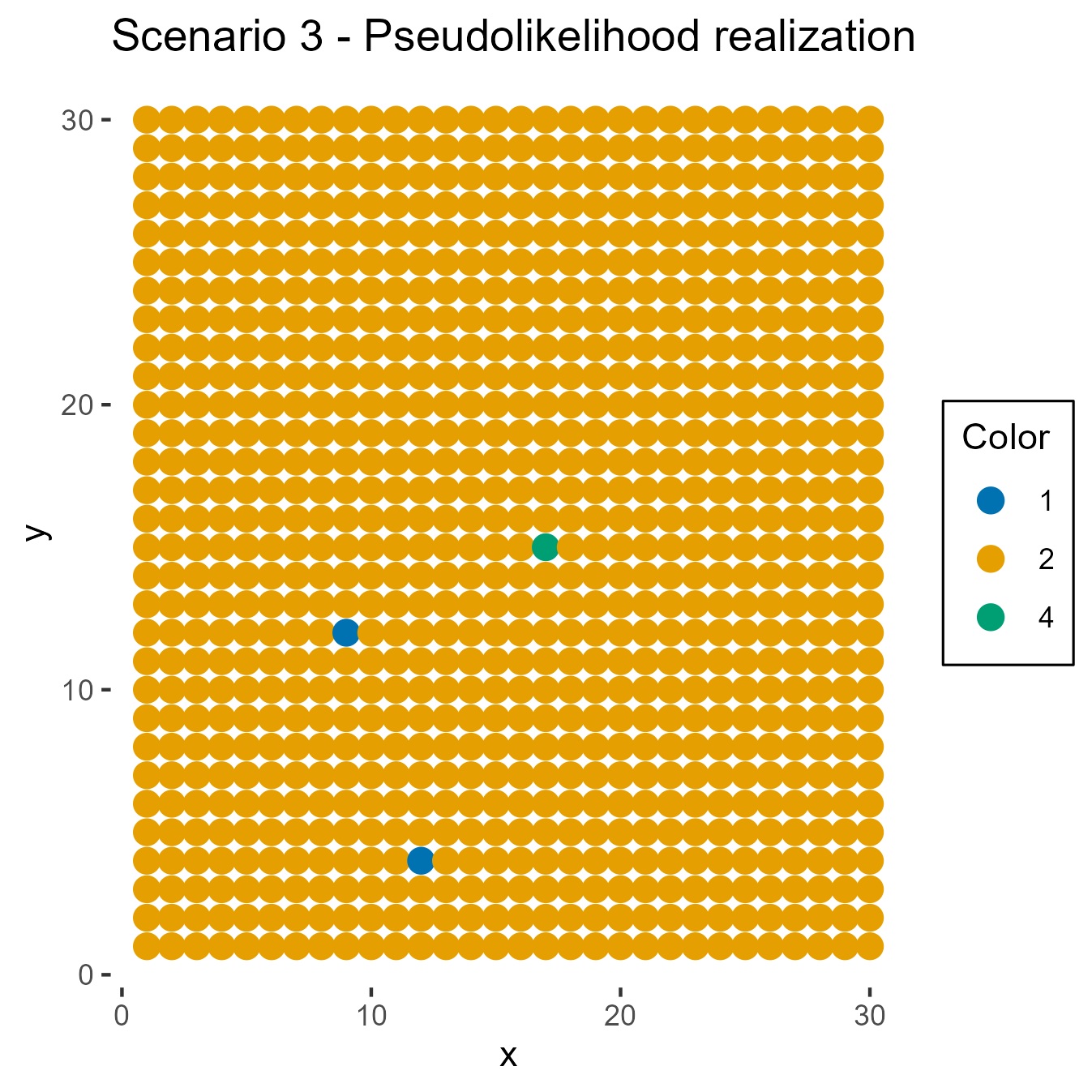}
    \includegraphics[width=5cm, height=3cm]{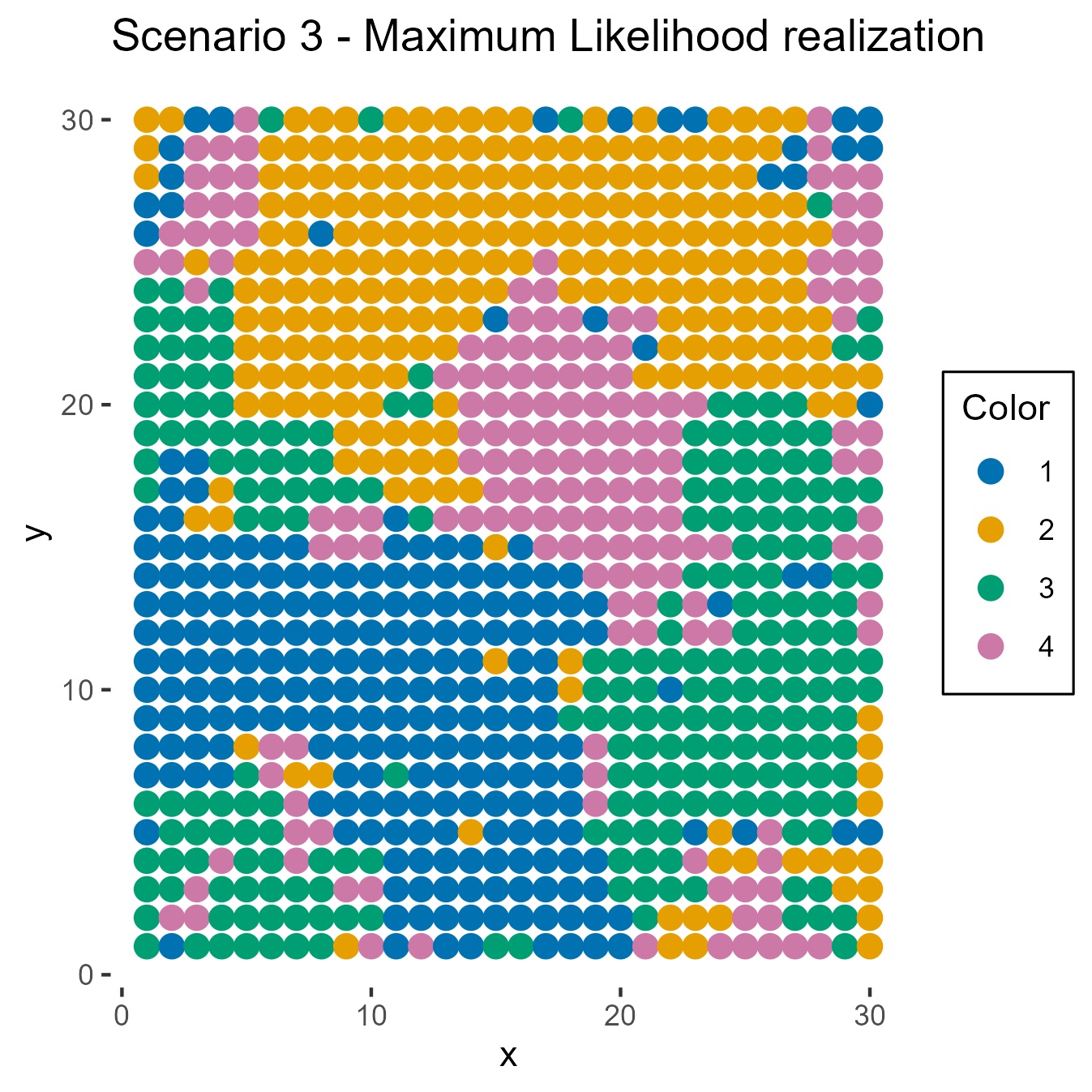}\\
    \includegraphics[width=5cm, height=3cm]{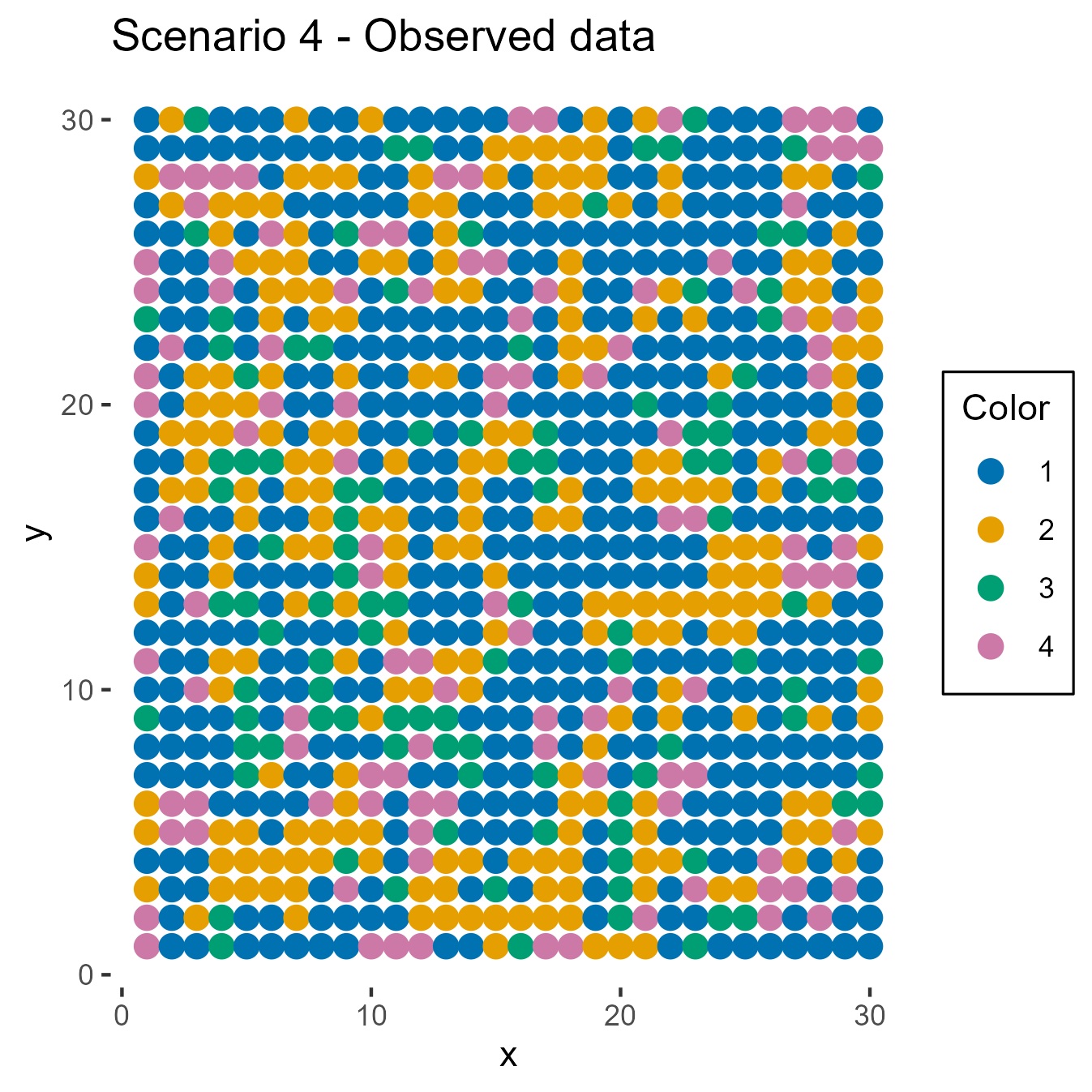}
    \includegraphics[width=5cm, height=3cm]{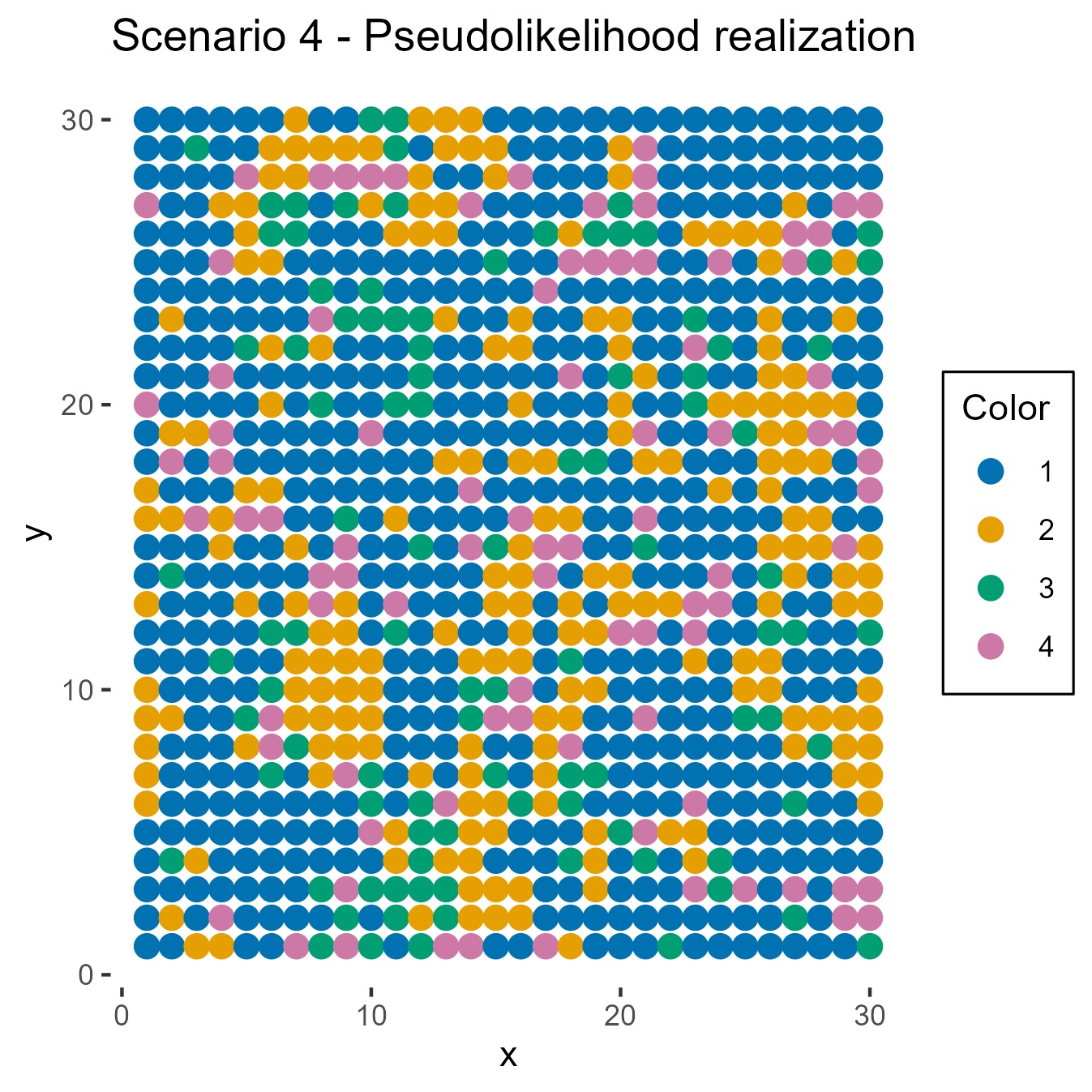}
    \includegraphics[width=5cm, height=3cm]{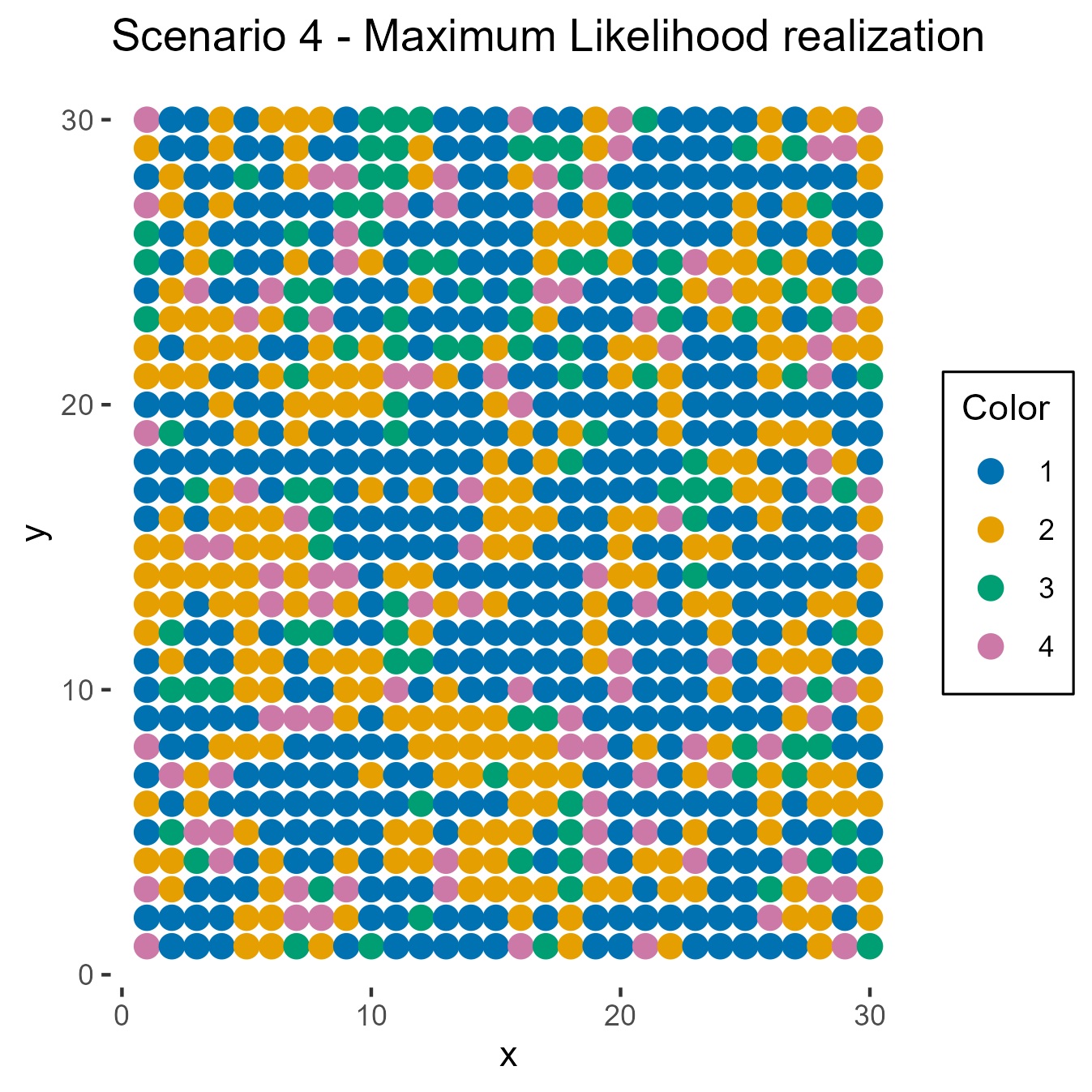}\\
     \includegraphics[width=5cm, height=3cm]{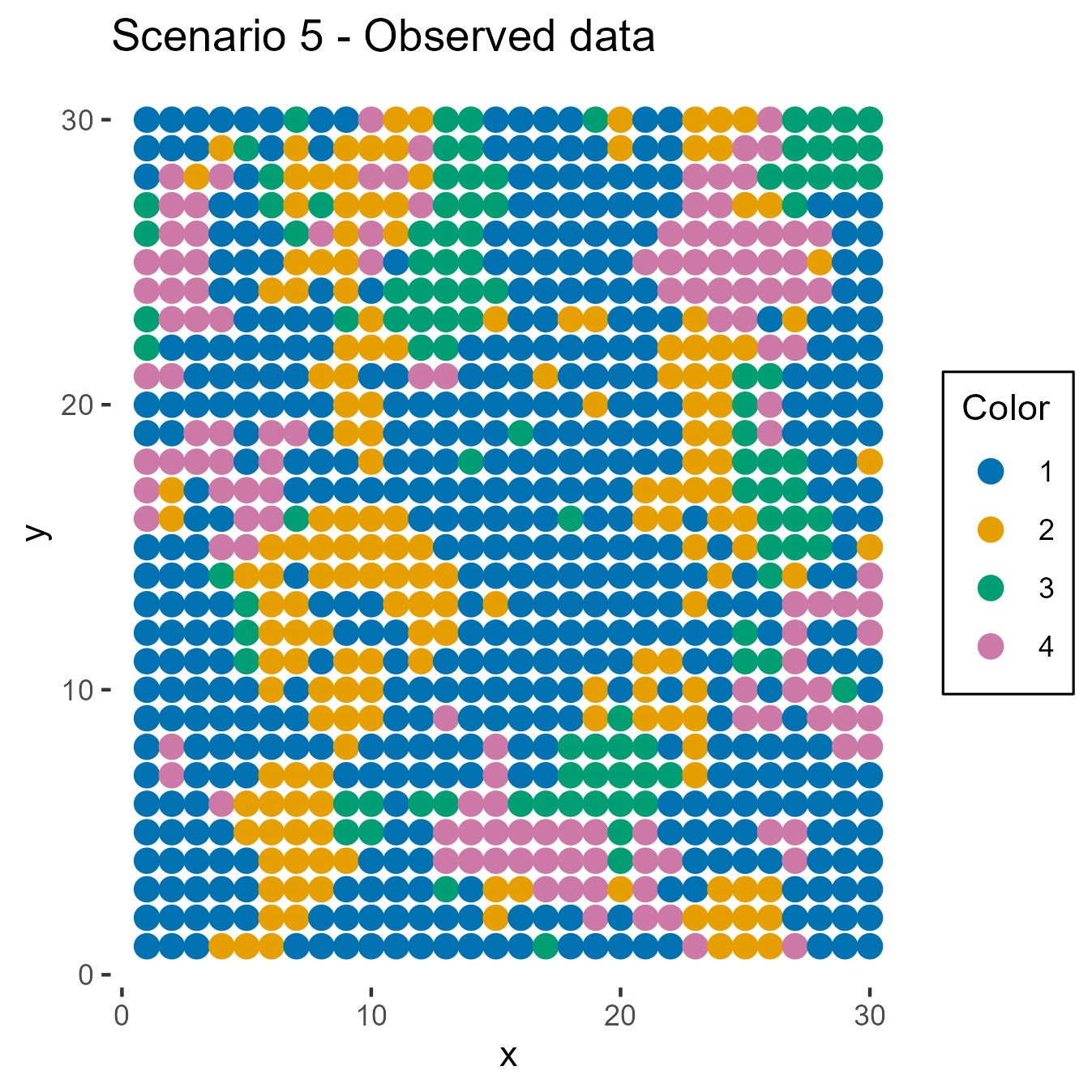}
    \includegraphics[width=5cm, height=3cm]{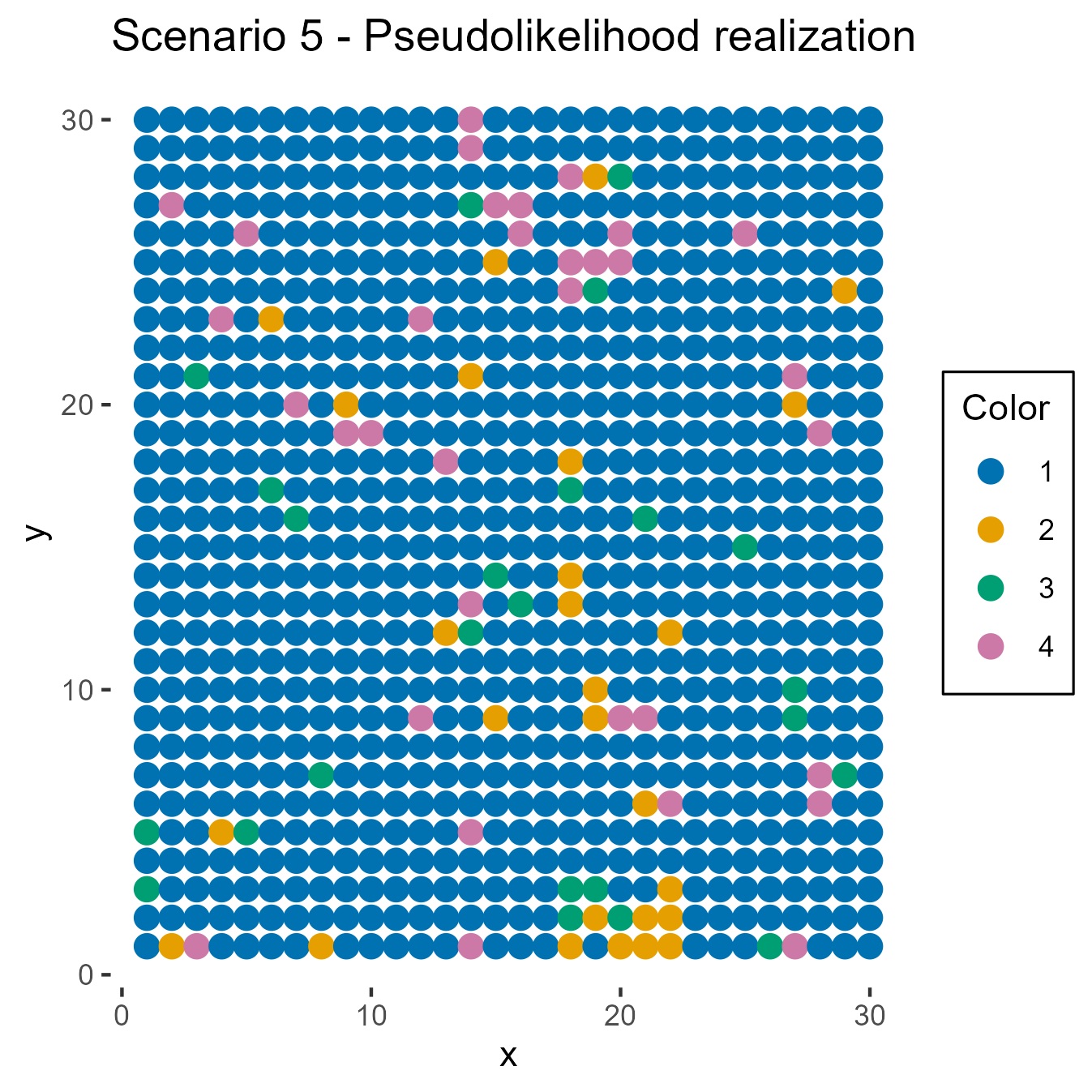}
    \includegraphics[width=5cm, height=3cm]{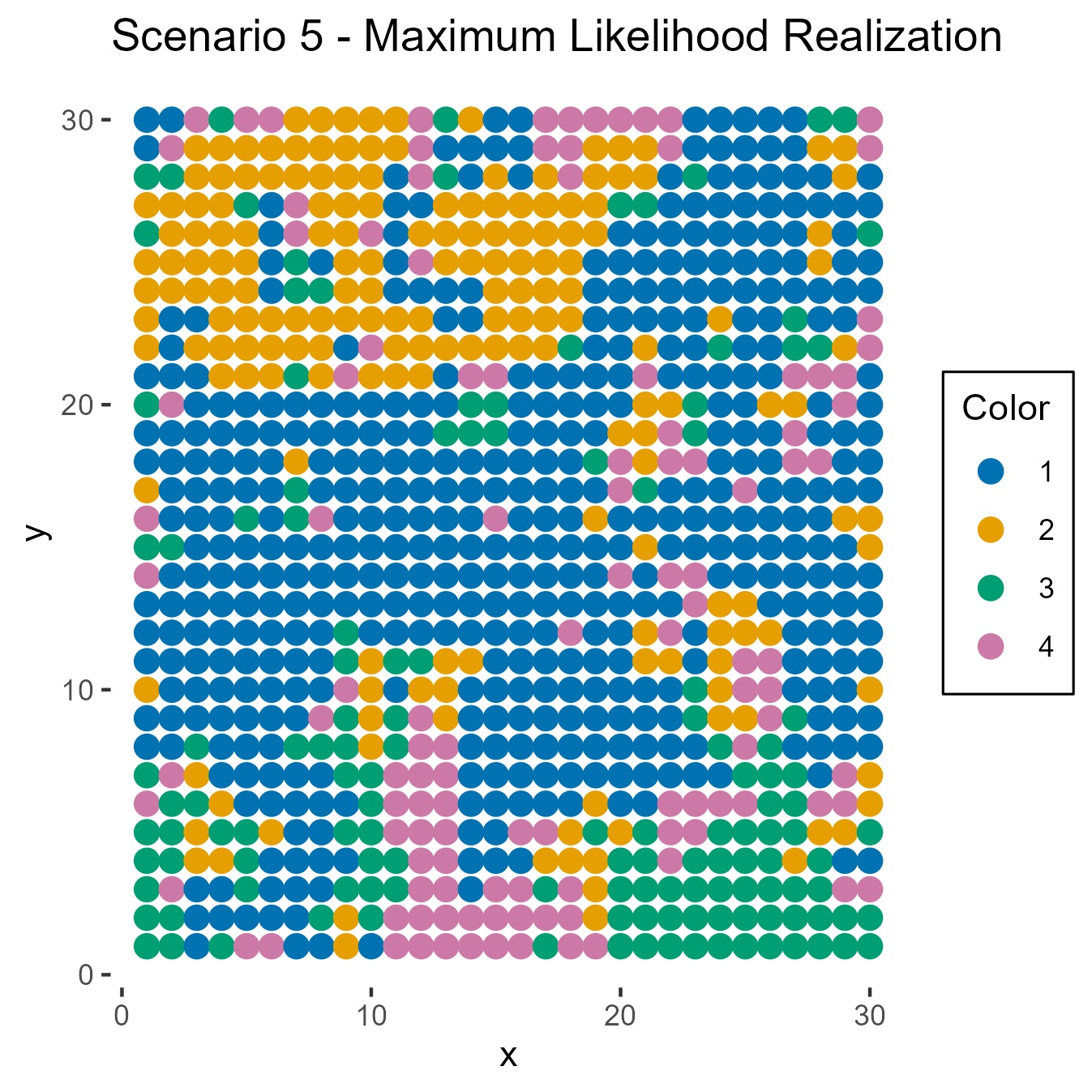}\\
     \includegraphics[width=5cm, height=3cm]{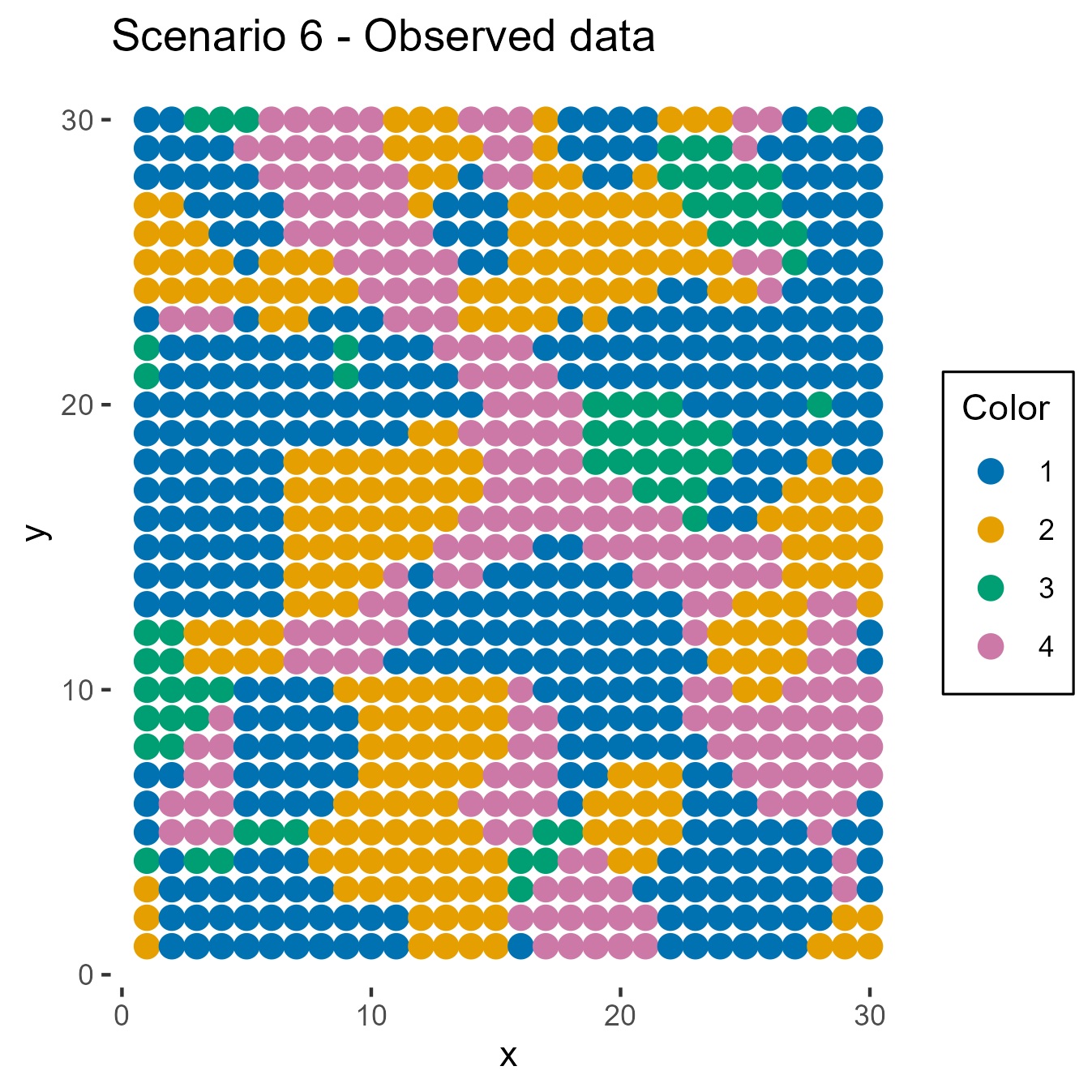}
    \includegraphics[width=5cm, height=3cm]{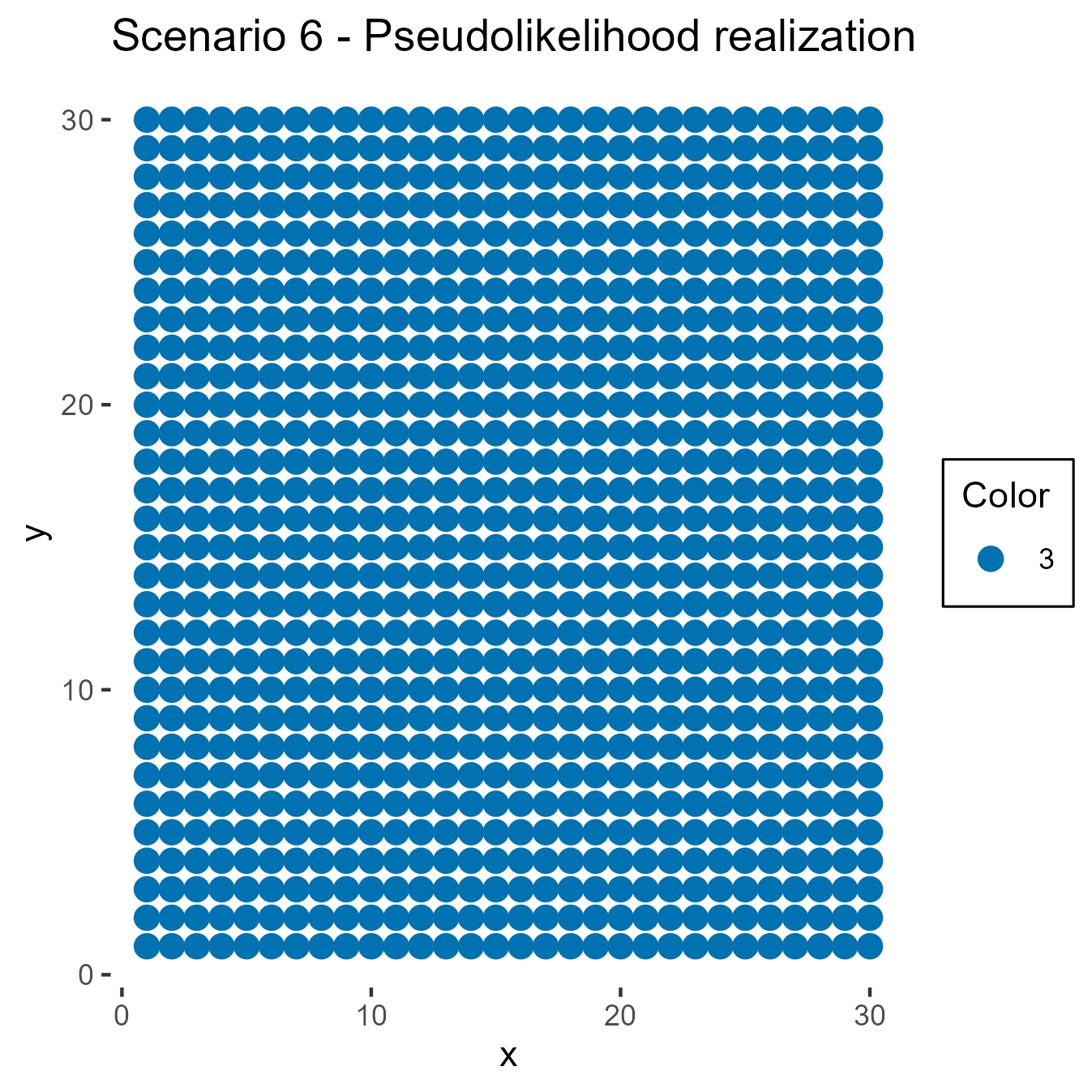}
    \includegraphics[width=5cm, height=3cm]{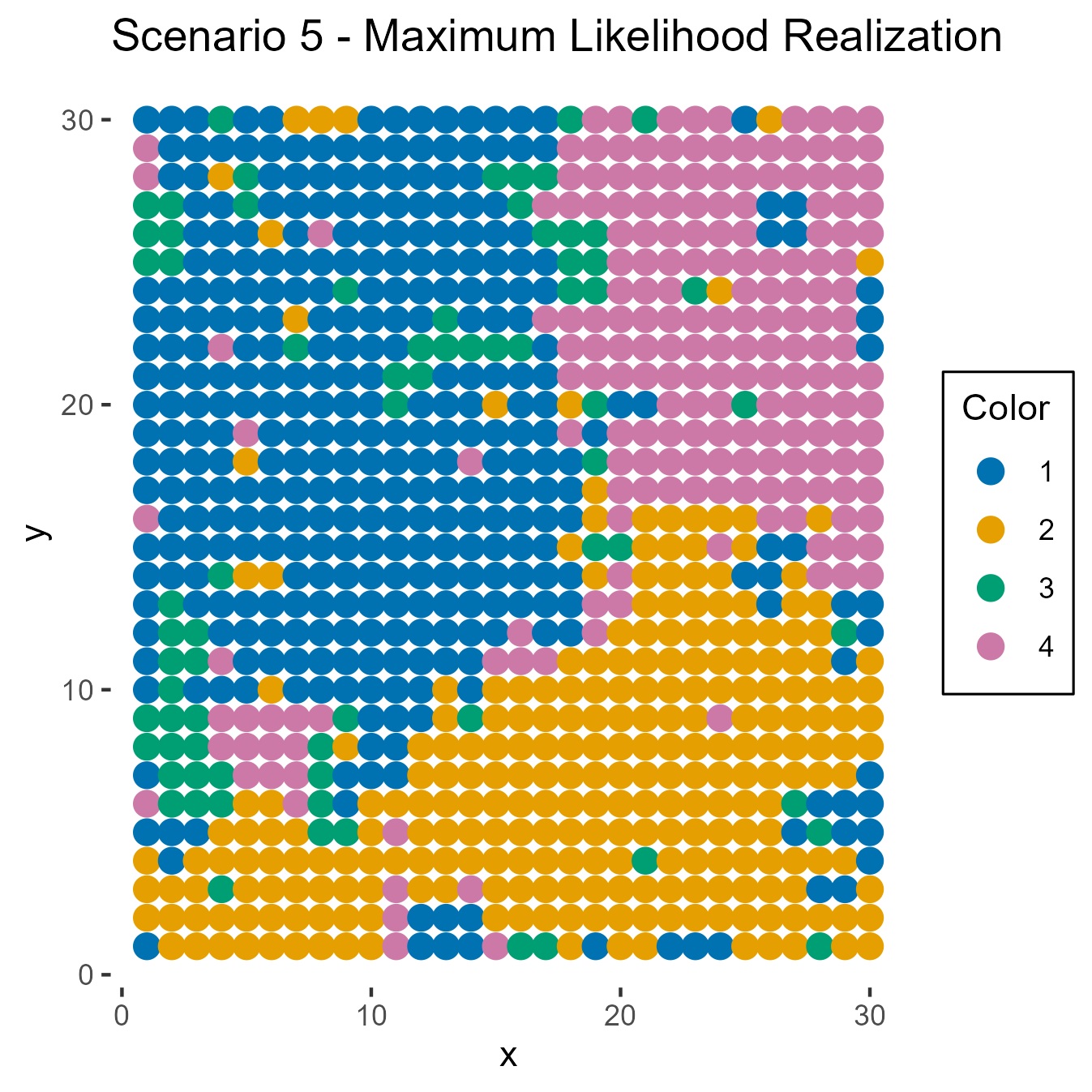}
    \caption{Original arrangements (left), realization from model fitted with pseudo-likelihood (center), and realization from model fitted with maximum-likelihood (right) for scenarios 1-6}
    \label{fig:Simulation_Prediction1}
\end{figure}

The resulting arrangements are presented in Figures \ref{fig:Simulation_Prediction1} (left).

For all six scenarios, we optimized the pseudo-likelihood \citep{Besag1974} to obtain initial values for the maximum-likelihood inference procedure. We attempted to fit a Potts model to the arrangements generated under all six scenarios. For scenarios 1 and 4, we successfully fitted a Potts model after a few iterations of the partial stepping algorithm. We found convergence problems when fitting the Potts model for scenarios 2, 3, 5 and 6. Therefore, we fitted the Tapered Potts model in those cases using the process described in Sections \ref{sec-MLTapPotts} and \ref{sec-ChooseTau}. Particularly, for scenario 2 and 3, since there is a similar proportion of sites of each color, we used the bimodality coefficient approach. The maximum-likelihood estimates of all the parameters are available in the supplemental materials. 

For each simulation scenario, we generated samples of 200 arrangements from the models that were fitted using pseudo-likelihood and maximum-likelihood. In Table \ref{tab:obssuff_predicted} we present the observed sufficient statistics associated with the original arrangement and the mean of the sufficient statistics from the fitted models in both cases.  For scenarios 1 and 4, where the pseudo-likelihood and maximum-likelihood estimates were relatively similar, we still found that the maximum-likelihood approach obtained sample means closer to the observed values in the original data. For the rest of the scenarios, where we had to fit a tapered Potts model, when comparing the sample means to the observed values we can see that the pseudo-likelihood fitted model does not provide good fit. In particular, it tends to choose one color and generate arrangements that are mostly of that color. The fitted pseudo-likelihood parameters also tend to generate arrangements with higher values of $S(\mathbf{x})$ compared to the observed data. In contrast, we found that the maximum-likelihood estimates generated sample means that are fairly close to the original values.

\begin{table}
    \centering
    \resizebox{\textwidth}{!}{\begin{tabular}{|c|c|c|c|c|c|c|}
    \hline
      Scenario & Method & $T_{1}^{obs}$ & $T_{2}^{obs}$ & $T_{3}^{obs}$ & $T_{4}^{obs}$ &  $S^{obs}$ \\
    \hline
       \multirow{4}{4em}{1}  & Observed & 210 &  219 &  236  & 235 &  687  \\
       & Pseudolikelihood Potts & 211.162 &  215.940 &  225.930  & 246.968 &   756.818  \\
       & Maximum likelihood Potts & 211.214 &  218.696 &  233.676  & 236.414 &  703.668  \\
       & Maximum likelihood Tapered Potts & NA &  NA &  NA  & NA &  NA \\
    \hline
    \multirow{4}{4em}{2}  & Observed & 225 &  178 &  165  & 332 &   1179  \\
       & Pseudolikelihood Potts& 815.876 &  28.676 &  27.190  & 28.258 &   1563.304  \\
       & Maximum likelihood Potts & NA &  NA &  NA  & NA & NA \\ 
       & Maximum likelihood  Tapered Potts& 224.800 &  177.865 &  164.945  & 332.390 &  1187.755  \\
    \hline
    \multirow{4}{4em}{3}  & Observed & 241 &  253 &  237  & 169 &   1297  \\
       & Pseudolikelihood Potts & 0.872 &  881.288 &  16.980  & 0.860 &   1790.094  \\
        & Maximum likelihood Potts & NA &  NA &  NA  & NA & NA \\ 
       & Maximum likelihood Tapered Potts & 240.925 &  253.115 &  236.910  & 169.050 &  1291.515  \\
    \hline
    \multirow{4}{4em}{4}  & Observed & 445 &  234 &  108  & 113 &   806  \\
       & Pseudolikelihood Potts & 473.258 &  216.626 &  103.048  & 107.068 &  882.580  \\
       & Maximum likelihood Potts & 450.914 &  234.054 &  106.284  & 108.748 &  836.402  \\
       & Maximum likelihood Tapered Potts & NA &  NA &  NA  & NA &  NA \\
    \hline
        \multirow{4}{4em}{5}  & Observed & 484 &  173 &  108  & 135 &   1143  \\
       & Pseudolikelihood Potts& 771.204 &  56.372 &  34.108  & 38.316 &  1469.776  \\
       & Maximum likelihood Potts & NA &  NA &  NA  & NA & NA \\ 
       & Maximum likelihood Tapered Potts & 496.120 &  171.815 &  100.505  & 131.560 &  1158.465  \\
    \hline
            \multirow{4}{4em}{6}  & Observed & 382 &  238 &  71  & 209 &   1315  \\
       & Pseudolikelihood & 0.876 &  1.022 &  897.174 & 0.928 &  1789.168 \\
       & Maximum likelihood Potts & NA &  NA &  NA  & NA & NA \\ 
       & Maximum likelihood Tapered Potts & 381.435 &  237.855 &  71.185  & 209.525 &  1332.81  \\
    \hline
    \end{tabular}}
    \caption{Observed sufficient statistics and mean sufficient statistics from samples generated from models fitted using pseudo-likelihood and Markov Chain Monte Carlo Maximum Likelihood for all simulation scenarios.}
    \label{tab:obssuff_predicted}
\end{table}

Figure \ref{fig:Simulation_Prediction1} compares the original realization (left) to a realization from the Potts model fitted with pseudo-likelihood (center), and a realization from the Potts/tapered Potts model fitted with maximum likelihood (right) in all simulation cases. For scenarios 1, 2, 4, and 5, we reproduced the properties of the original arrangements using maximum-likelihood. However, in scenarios 3 and 6, although we reproduced arrangements with similar observed sufficient statistics, the original arrangements present less isolated points compared to the observations generated from the fitted models. Based on this simulation study, we conclude that the Potts and tapered Potts successfully reproduced the characteristics of a variety of multi-category data on a lattice. However, based on the results from scenarios 3 and 6, we note that the Tapered Potts may not reproduce highly smooth arrangements with little noise.

\section{Application to National Land Cover Database data}\label{sec-NLCDIntro}

The National Land Cover Database (NLCD) provides nationwide data on land cover at the Landsat Thematic Mapper (TM) 30-meter resolution \citep{NLCD}. In this work, we consider the information from 2021, choosing two small areas of the US that were partitioned into a regular grid: a small area of the state of Pennsylvania (longitude: $-75.2$ to $-75$, latitude: $41.72$ to $42$) and the Mesa Verde National Park in Colorado. The main objective of this application is to reproduce the land cover composition and level of spatial correlation in the areas of interest. We found that the traditional Potts model fits the Pennsylvania data well, while it fits poorly to the Colorado data. The database was accessed through the \textit{FedData} R package \citep{FedData}.

\subsection{Subregion of Pennsylvania}\label{sec-Pennsylvania}

\begin{figure}
        \centering
        \includegraphics[width=3.5cm, height=5cm]{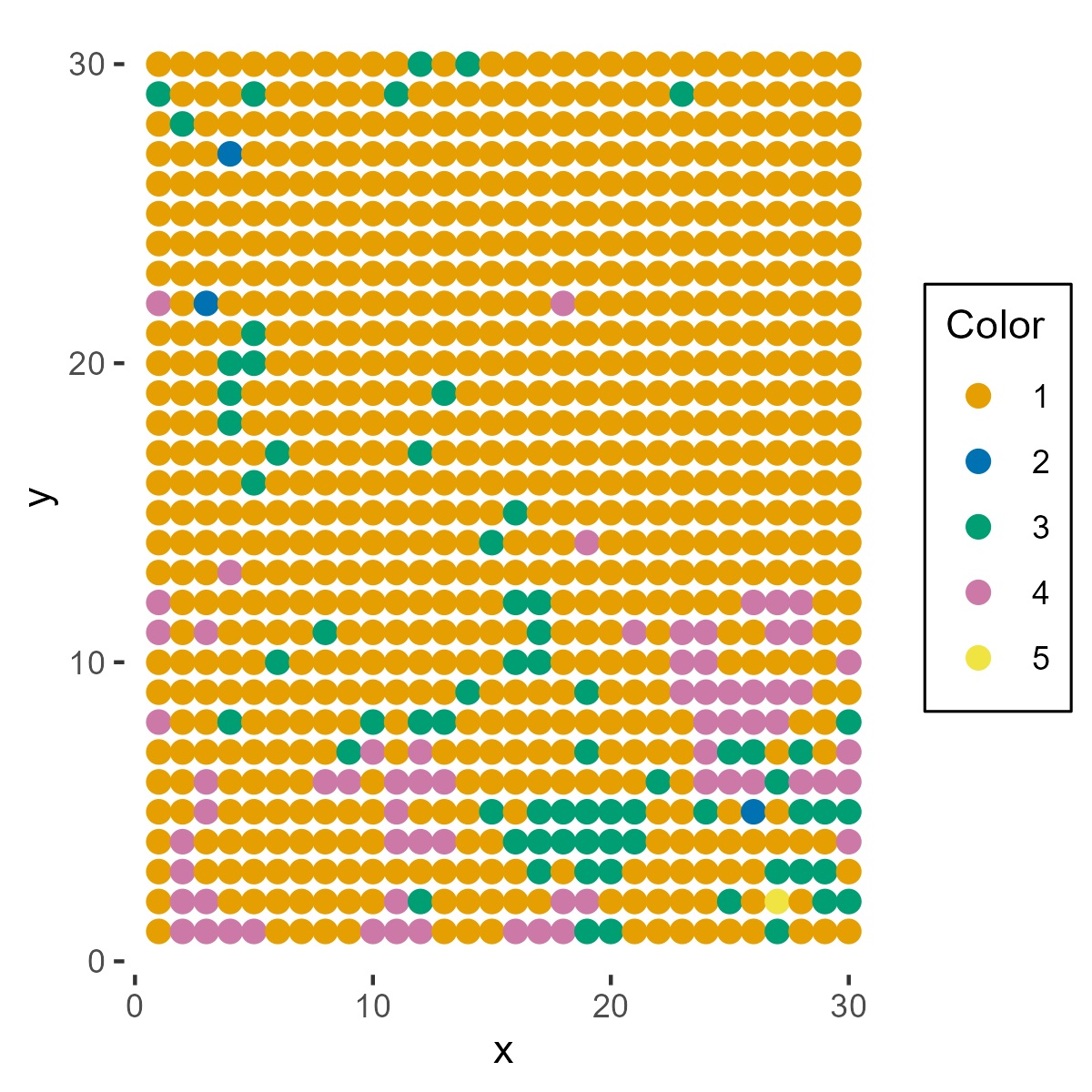}
        \includegraphics[width=3.5cm, height=5cm]{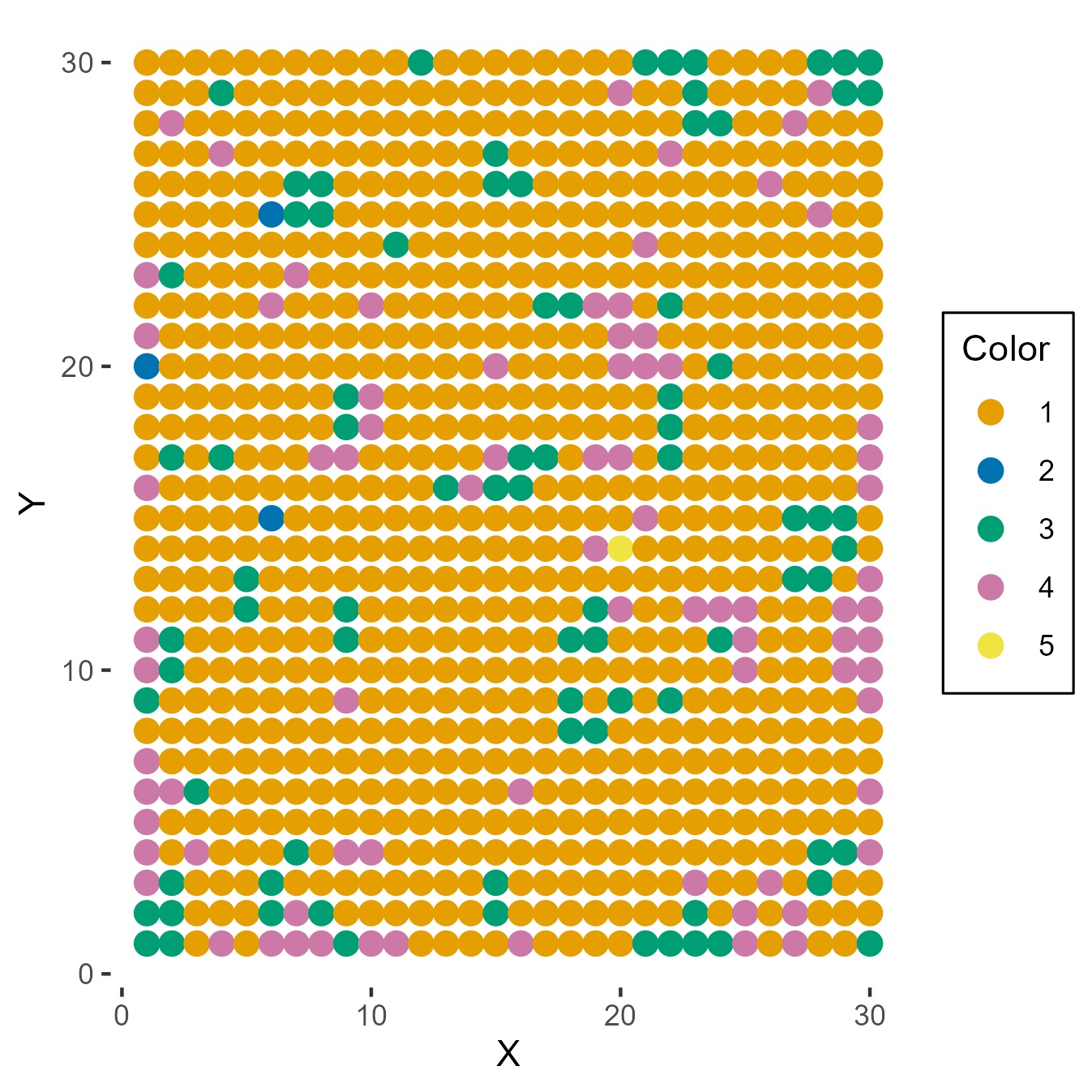}
        \includegraphics[width=3.5cm, height=5cm]{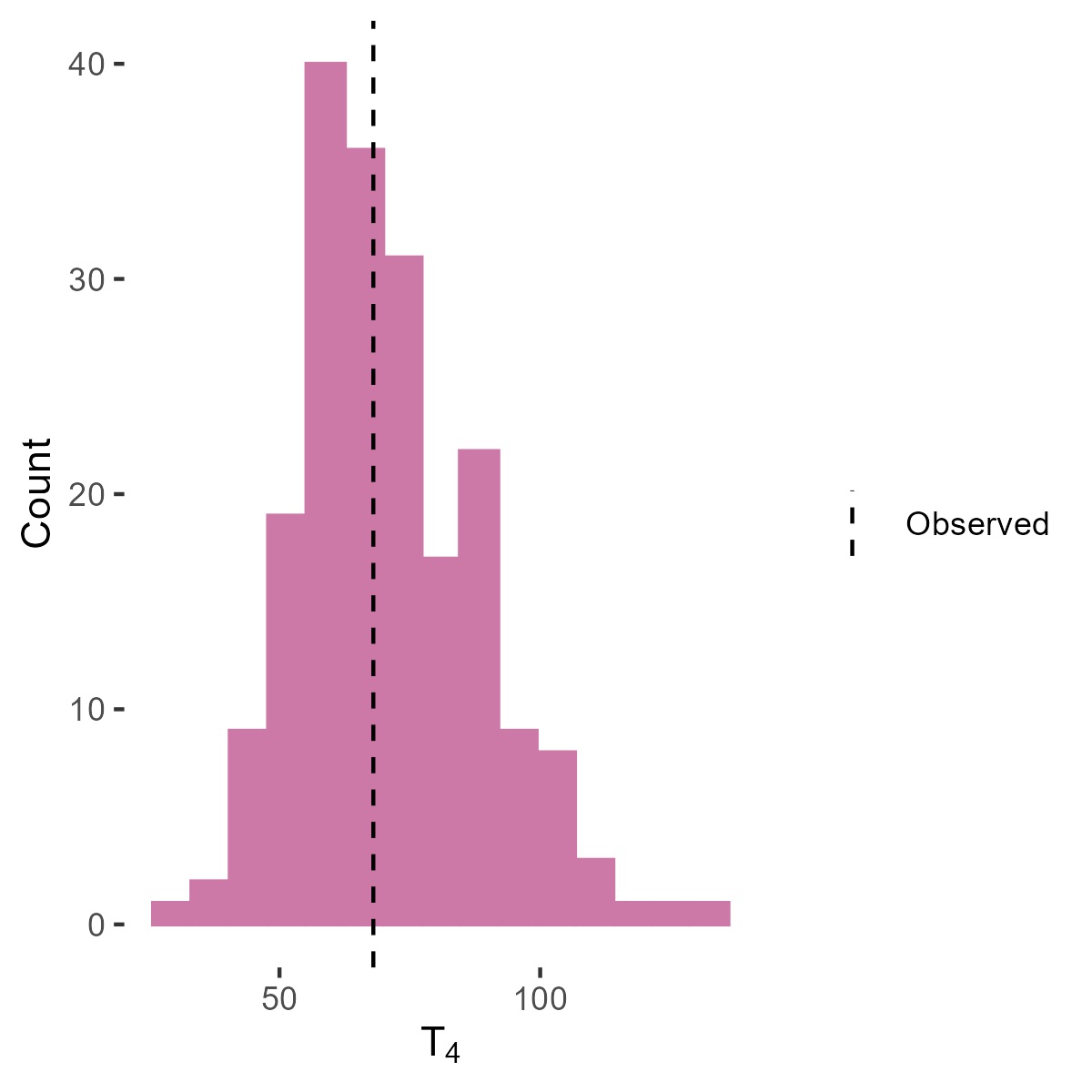}
        \caption{Observed data in a subregion of Pennsylvania (left), arrangement generated by the approximate maximum likelihood estimates (center), and sample distribution of $T_{4}(\mathbf{x})$ using the approximate maximum-likelihood estimates (right).}
        \label{fig:MapMLEPenn}
    \end{figure}

After partitioning the original area of interest into a $30 \times 30$ regular grid, we assigned the predominant land cover to each cell. The land cover types present in the arrangement after this process are:  Deciduous Forest (color 1), Developed, Open Space (color 2), Mixed Forest (color 3), Pasture/Hay (color 4), and Woody Wetlands (color 5). The observed arrangement is presented in Figure \ref{fig:MapMLEPenn} (left).

Because we have multi-category data on a lattice with evident spatial clustering, the Potts model is a plausible option for modeling. According to the observed sufficient statistics, type 1 is the predominant land cover class in the arrangement. However, there is an important presence of other land cover types, such as types 3 and 4. Thus, it is necessary to determine whether the level of spatial dependence present in the data is high enough to cause a lack of fit of the Potts model.

To obtain initial values for the partial stepping algorithm and gain a preliminary understanding of the parameter values, we maximize the pseudo-likelihood of the Potts model for this dataset. Based on the pseudo-likelihood estimate of $\hat{\beta}_{PL}=0.801$, it is likely that the Potts model is a good fit for this arrangement. After only six iterations of the partial stepping algorithm, we found a value of $\boldsymbol{\theta}_0$ capable of reproducing a sample mean close to $\mathbf{T}(\mathbf{x^{obs}})$. This indicates that the Potts model provides a reasonable fit for the observed arrangement.

Figure \ref{fig:MapMLEPenn} (center) presents an arrangement generated from the fitted Potts model using the approximate maximum-likelihood estimates, and Figure \ref{fig:MapMLEPenn} (right) presents the empirical distribution of $T_{4}(\mathbf{x})$ based on a sample of 500 arrangements from the fitted model. We conclude that the Potts model effectively reproduces the land cover composition of this region of Pennsylvania, preserving both the approximate proportions of each cell type and the spatial clustering observed in the original dataset.

\subsection{Mesa Verde National Park}

\begin{figure}
    \centering
    \includegraphics[width=3.5cm, height=5cm]{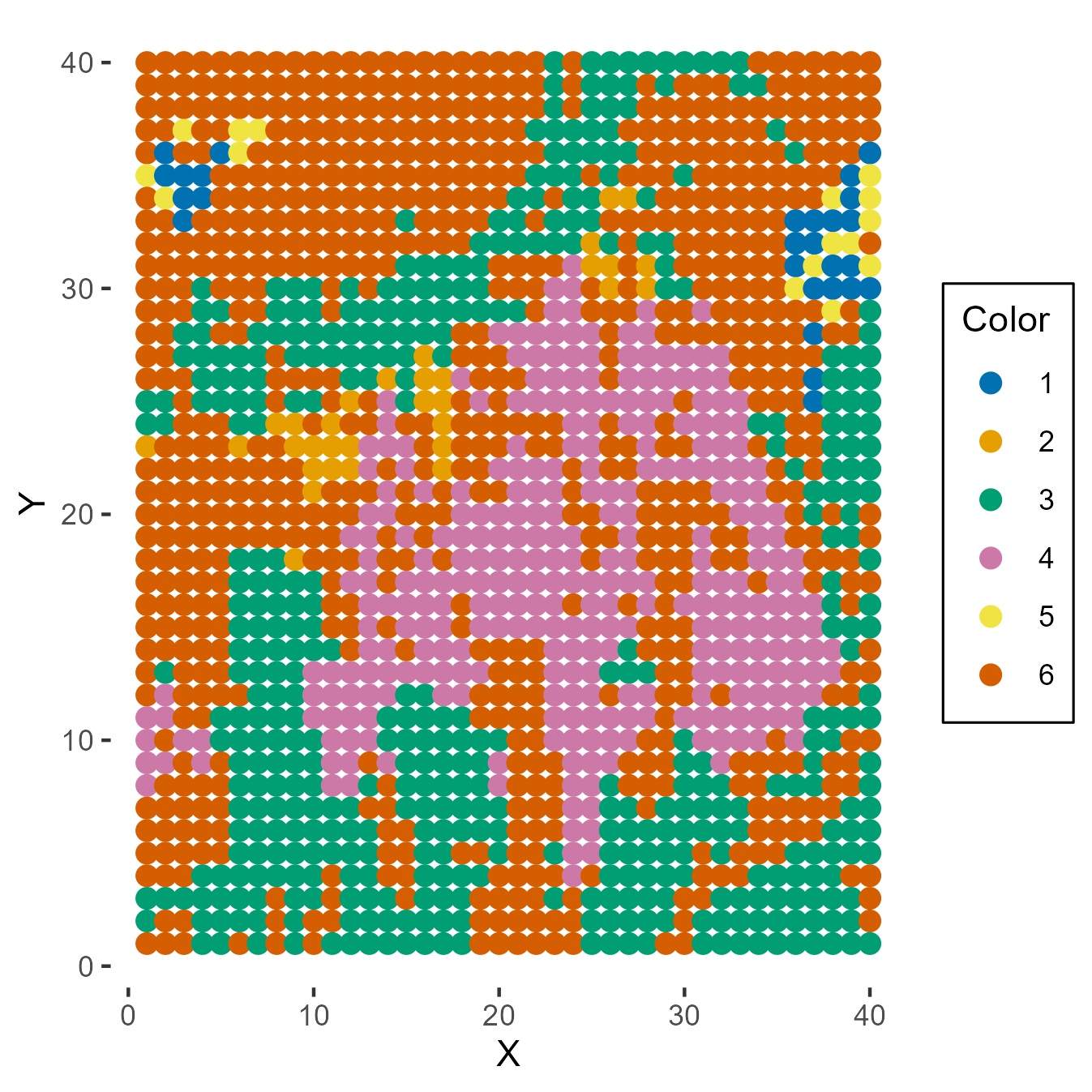}
    \includegraphics[width=3.5cm, height=5cm]{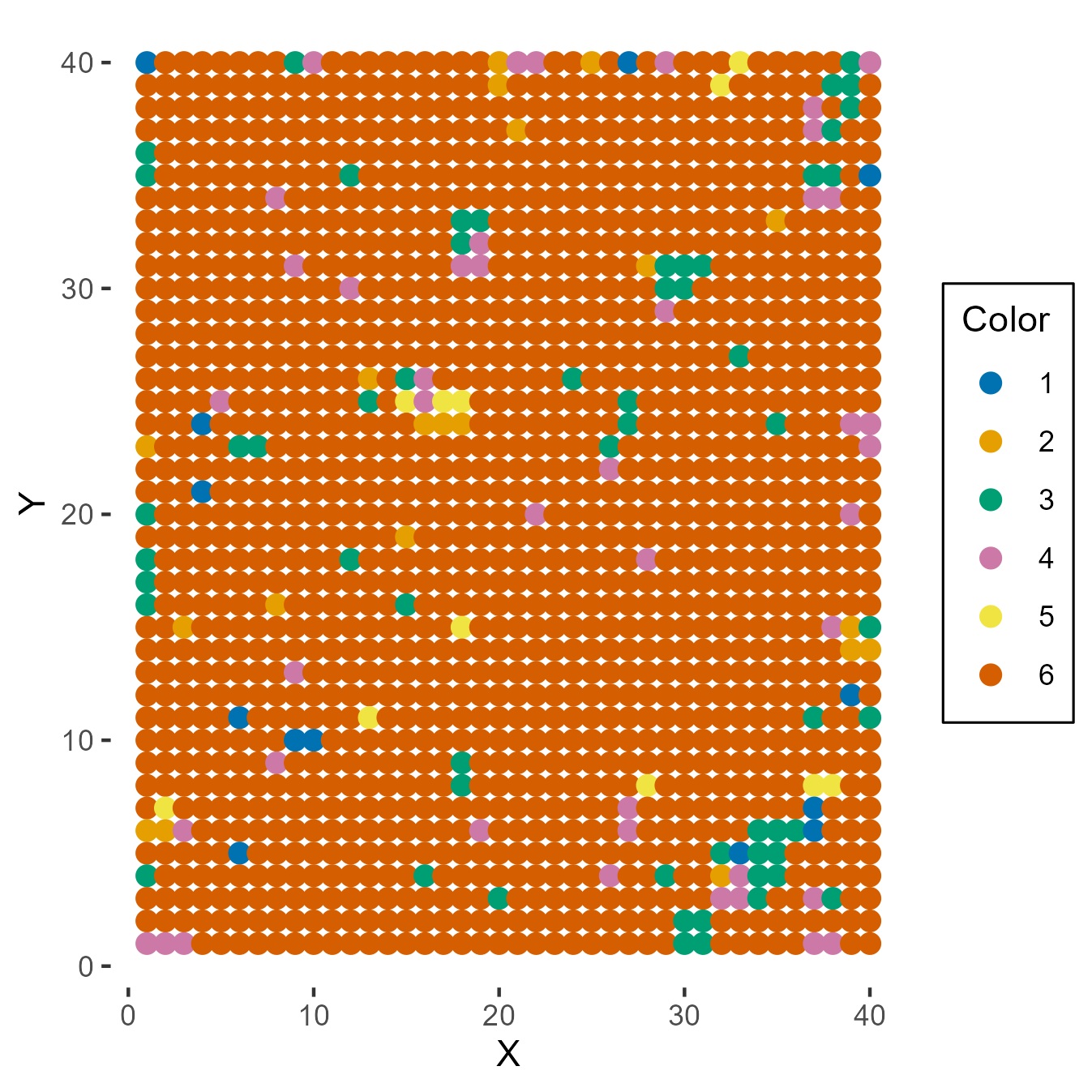}
    \includegraphics[width=3.5cm, height=5cm]{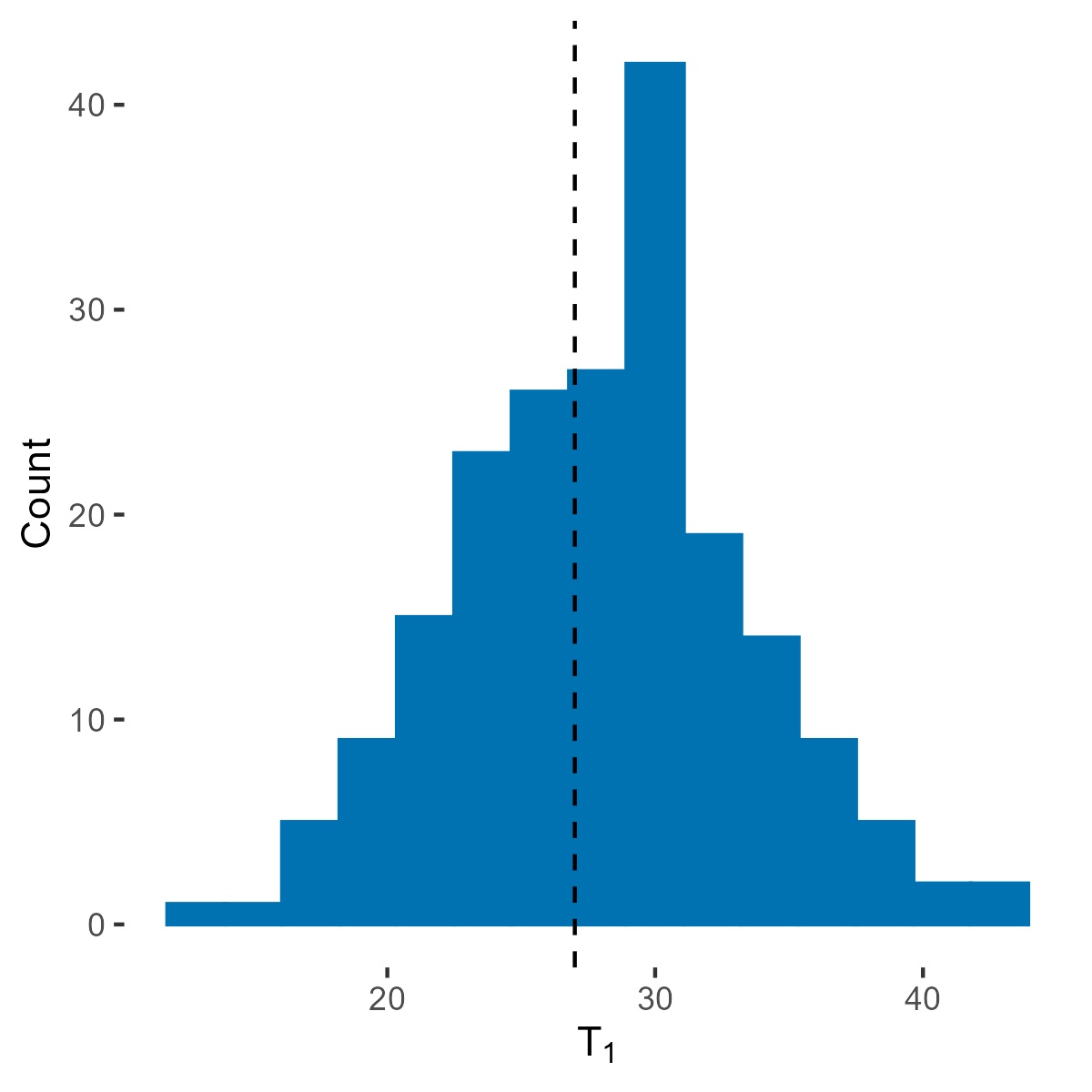}
    \includegraphics[width=3.5cm, height=5cm]{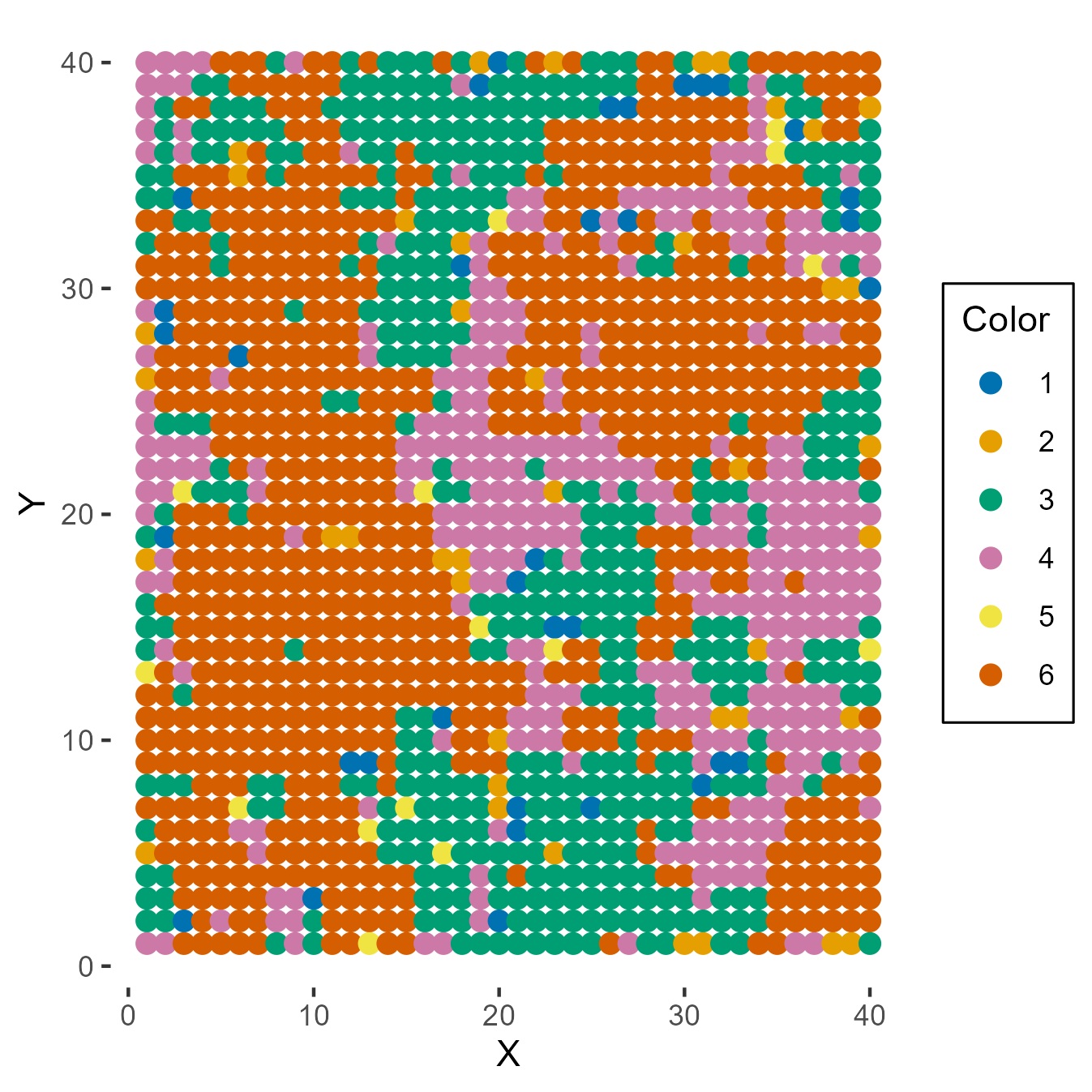}
    \caption{Observed arrangement for Mesa Verde National Park (a), 
    arrangement generated by fitted Potts model using pseudo-likelihood (b), sample distribution of $T_{1}(\mathbf{x})$ using the approximate maximum-likelihood estimates of the Tapered Potts model (c), and arrangement generated by the fitted Tapered Potts model (d).}
    \label{fig:MVNPGrid}
\end{figure}

We partitioned the original Mesa Verde National Park area into a $40 \times 40$ regular grid, and we assigned the predominant land cover to each cell. After this procedure, we maintain a total of six predominant land cover types: Cultivated Crops (color 1), Deciduous Forest (color 2), Evergreen Forest (color 3), Grassland/Herbaceous (color 4), Pasture/Hay (color 5), Shrub/Scrub (color 6). The resulting arrangement is presented in Figure \ref{fig:MVNPGrid} (a).

Type 6, which is the reference category, is the predominant land cover type in this dataset. We maximized the pseudo-likelihood of the Potts model for this dataset and obtained $\hat{\beta}_{PL}=1.098$, which suggests that there is a high level of spatial correlation in the data and it may be necessary to include a tapering term. Figure \ref{fig:MVNPGrid} (b) presents an arrangement generated using the Potts model parametrized by the pseudo-likelihood estimates. Next, we apply the practical guidelines outlined in Section \ref{sec-LackofFit} by using the partial stepping algorithm to assess whether it is necessary to introduce a tapering term. We observed that the average number of sites of color 1 decreases to almost zero as the algorithm progresses. We also noted that the partial stepping algorithm oscillates between nearly complete arrangements of color 3 and color 6 during the first 8 iterations. When color 3 dominates, color 6 is almost absent, and vice versa. These oscillations indicate that the algorithm is not converging, a consequence of the ground states inherent in the Potts model under such a high level of spatial correlation. This confirms that the Potts model cannot reproduce the composition the Mesa Verde National Park land cover arrangement and that we need to add a tapering term.

From all of our simulation studies, we did not find evidence that the tapering values should be different for each statistic $T_{k}(\mathbf{x})$. Therefore, we set $\tau_k=\tau$ for all $k=1, \ldots, K-1$. In this example, we initialize the algorithm with $\tau=0.002$, a relatively large value in this context, to ensure convergence in the first step. We then gradually reduce the tapering parameter in subsequent steps. After 10 iterations, decreasing $\tau$ by $10\%$ at each iteration, we found that the smallest $\tau$ ensuring convergence was $0.0001917731$. Using this value, the approximate maximum likelihood estimates were found.

Figure \ref{fig:MVNPGrid} (c) shows the sample distributions of $T_{1}(\mathbf{x})$ based on 200 simulated arrangements from the fitted tapered Potts model. The distribution is centered around the observed values in the data, indicated by the dashed lines. The behavior was also observed in the empirical distributions of the statistics $T_{j}(\mathbf{x})$. Additionally, Figure \ref{fig:MVNPGrid} (d) shows a representative simulated arrangement that closely replicates the original land cover composition and its spatial clustering.

\section{Discussion}\label{sec-Discussion}

In this work, we have studied the behavior of the Potts and Ising models in various simulated and real data scenarios. In particular, we studied their characteristics in the presence of an external field and examined their behavior under phase transition and described their ground states across different scenarios. We also illustrated the computational challenges and lack-of-fit issues arising from these behaviors. Furthermore, we provided a comprehensive guide for performing inference with the Potts model, along with practical advice for identifying situations where the model exhibits lack of fit and may be inappropriate. To address such cases, we proposed a modification called the tapered Potts model. We applied the proposed methodology to both a simulation study and a land cover dataset. Our results demonstrate that the Potts and tapered Potts models are effective tools for modeling discrete lattice data, accommodating diverse spatial patterns. These models are particularly well-suited for applications such as clustering and analyzing lattice compositions with multinomial responses, offering enhanced flexibility to capture a broad range of configurations. The guidelines developed in this study provide a practical framework for implementing the tapered Potts model on real-world datasets.

\section{Acknowledgments}
We thank John Hughes and Ephraim Hanks for their insightful comments. 

\section{Disclosure Statement}
The authors report there are no competing interests to declare.

\newpage

\section{Supplementary Material for ``Modeling discrete lattice data using the Potts and tapered Potts models''}

Data availability: The data that support the findings of this study are publicly available in the US Geological Survey's National Land Cover Database at \url{https://www.usgs.gov/centers/eros/science/national-land-cover-database} .

\subsection{Realizations of the Potts model}

\begin{figure}
    \centering
    \includegraphics[width=6cm, height=6cm]{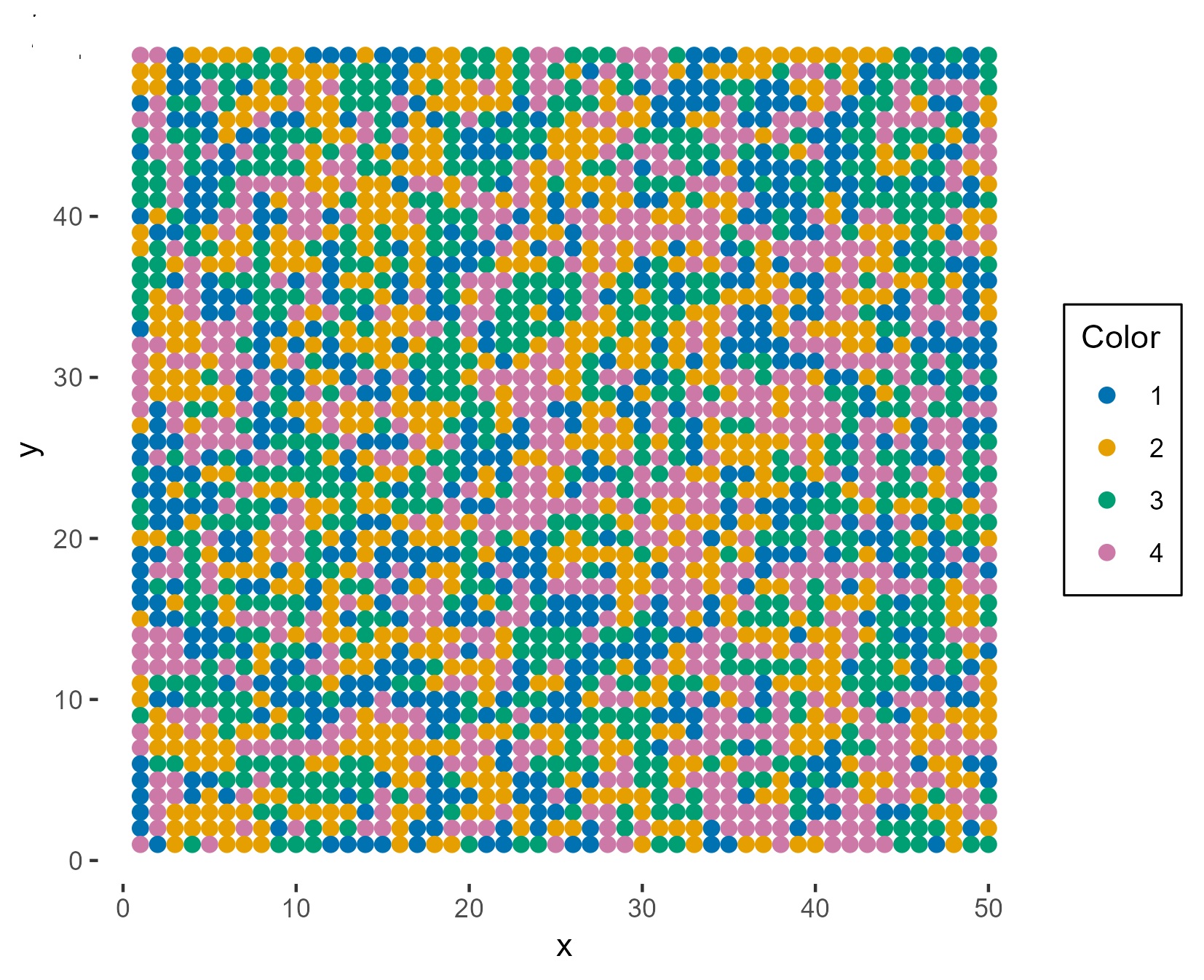}
    \includegraphics[width=6cm, height=6cm]{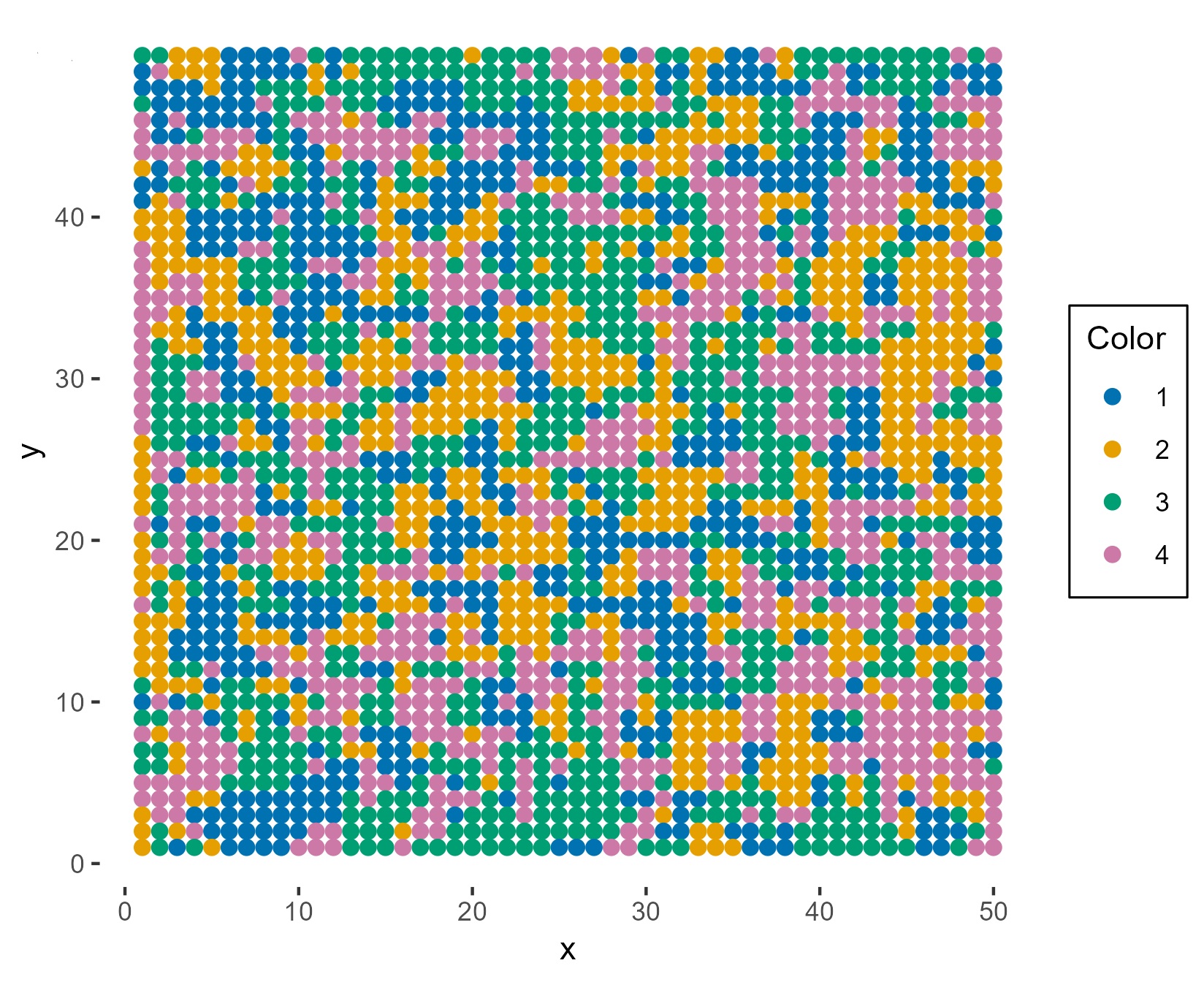}
    \caption{Realizations from Potts model with  $\boldsymbol{\alpha}=(0,0,0)^t$, and (left) $\beta=0.5$ and (right) $\beta=0.9$.}
    \label{fig:ExPotts1}
\end{figure}

Figure \ref{fig:ExPotts1} presents two realizations from the Potts model with $K=4$ on a lattice of size $50 \times 50$. Both configurations were generated using $\boldsymbol{\alpha}=(0,0,0)^t$, which means that we expect to observe similar proportions of cells of each color. However, we observe larger clusters of colors in Figure \ref{fig:ExPotts1} (b), where there is higher spatial correlation. 

\begin{figure}
    \centering
    \includegraphics[width=6cm, height=6cm]{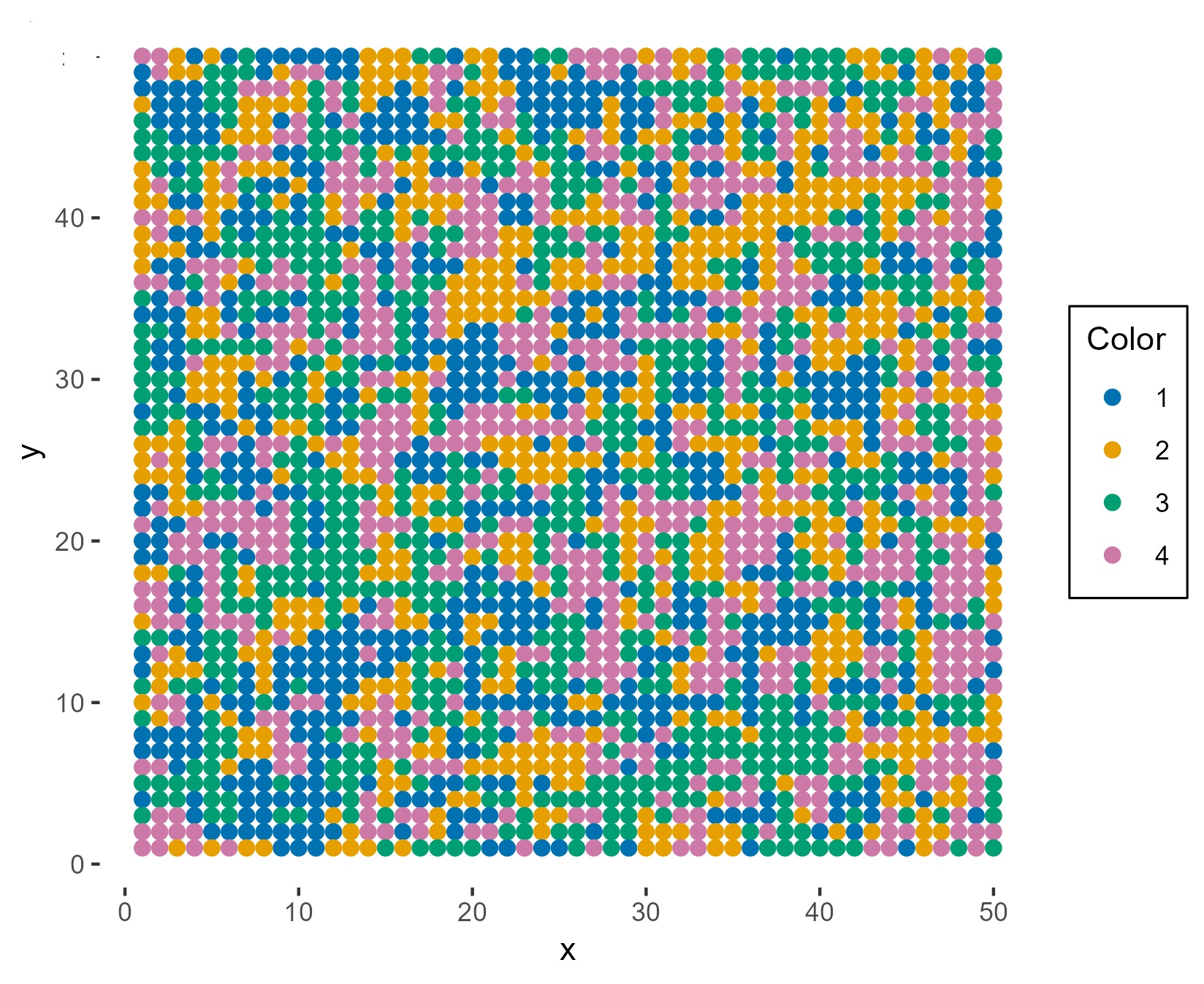}
    \includegraphics[width=6cm, height=6cm]{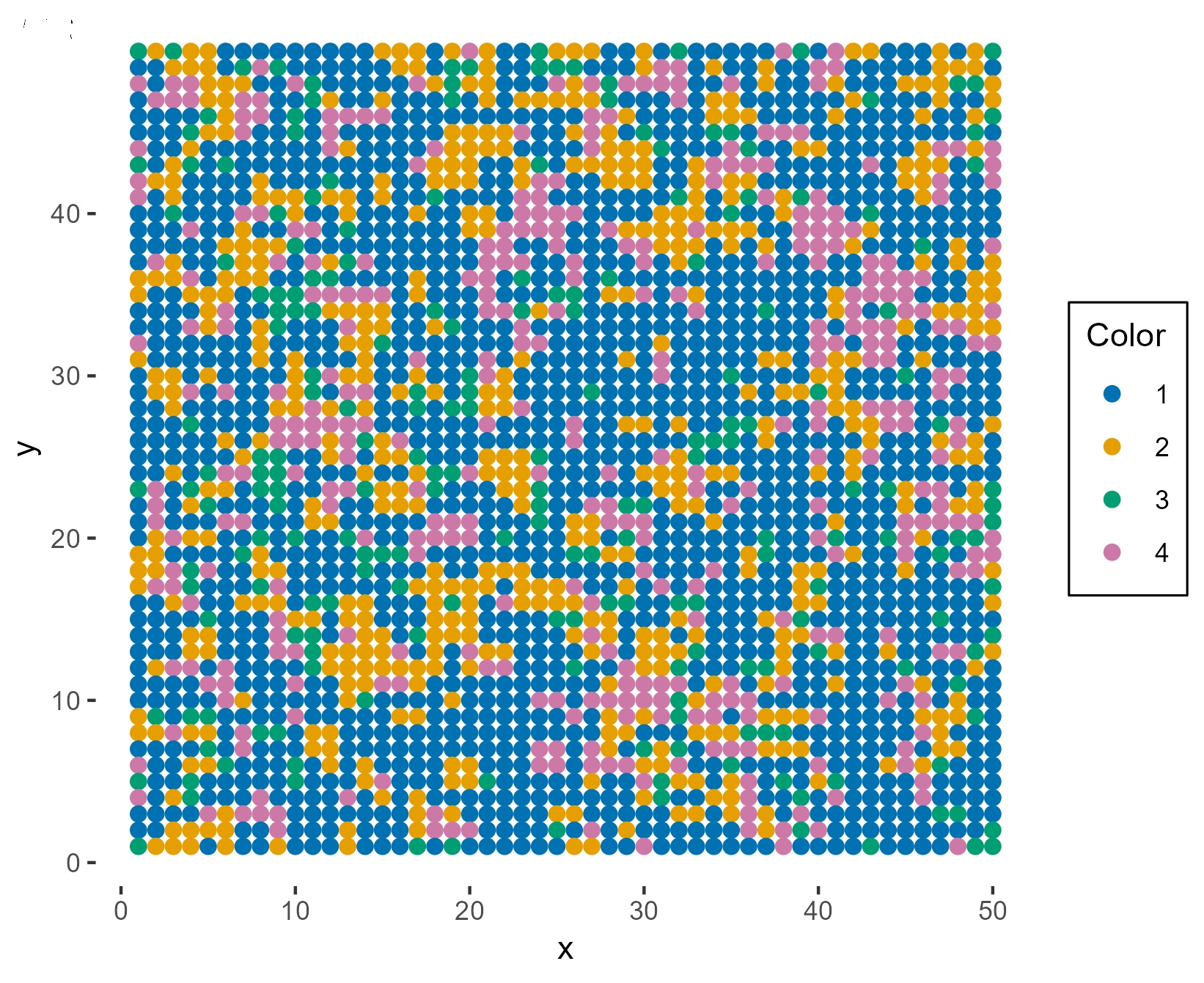}
    \caption{Realizations from Potts model with $\beta=0.7$, and (left) $\boldsymbol{\alpha}=(0,0,0)^t$ and (right) $\boldsymbol{\alpha}=(0.3,0.1,-0.3)^t$.}
    \label{fig:ExPotts2}
\end{figure}

On the other hand, Figure \ref{fig:ExPotts2} presents two realizations of the Potts model with the same level of spatial dependence ($\beta=0.7$) but different values of $\boldsymbol{\alpha}$. In Figure  \ref{fig:ExPotts2}  (a),  we expect to observe similar proportions of cells of each color. In Figure  \ref{fig:ExPotts2}  (b), we expect to observe a higher proportion of cells of colors 1 and 2 compared to the proportion of cells of color 4, and a smaller proportion of cells of color 3 compared to color 4. 

\subsection{Gibbs sampler for the classical Potts model} \label{sec-GibbsPotts}

For a lattice of size $M = n \times m$,

\begin{enumerate}
    \item Fix the values for the parameters $\boldsymbol{\alpha}$ and $\beta$.
    \item Randomly select one color $k \in \{1,\ldots,K\}$. Let $\mathbf{x}$ be an initial value of the observed arrangement for the sampler where all the realizations $x_{i}=k$ for $i=1,\ldots, M$.
    \item Generate a list of neighbors or a contiguity matrix. This is necessary to calculate the change of the statistic $S$ at each iteration. 
    \item For a site $i$, randomly select a value $l$ from the set $\{1,\ldots,k-1, k+1, \ldots, K\}$. Let $x_{i}$ be the current value of the arrangement at site $i$, and calculate
$$r=\text{min}\biggl\{1, \exp\bigl\{\alpha_l-\alpha_k + \beta \sum_{i \sim j} I\left(l=x_{j}\right) - I\left(x_{i}=x_{j}\right)\bigl\} \biggl\}.$$
Let $U \sim \text{Uniform}(0,1)$. If $u<r$, then set $x_{i}$. Otherwise, keep the previous value $x_{i}$. Repeat this step for all locations $i=1,\ldots, M$ to complete a full simulation cycle.

\item Once step 3 is completed, attempt to swap the values of the arrangement as follows:
\begin{enumerate}
    \item Generate a random ordering of the values $\{1,\ldots, K\}$ denoted by $\{r_1,\ldots, r_K\}$. We will create a new arrangement $\mathbf{y}$ where for every site $j$ such that if $x_{j}=k$, then $y_{j}=r_k$, $k=1,\ldots, K$.
    \item Note that $S(\mathbf{y})=S(\mathbf{x})$ and $T_{r_k}(\mathbf{y})=T_{k}(\mathbf{x})$. Let $\mathbf{T}(\mathbf{x})=\left(T_1(\mathbf{x}), \ldots, T_K(\mathbf{x})\right)^t$, and calculate the odds
    $$odds=\exp\Bigl\{ \alpha^{t}\left(T(\mathbf{y})-T(\mathbf{x}) \right)\Bigl\}$$
\end{enumerate}
If $odds \geq 1$, $\mathbf{y}$ is set as a new observation of the model. Otherwise, $\mathbf{x}$ will be the new element of the sample.
    \item Repeat steps 4 and 5 a total of $s+b$ times, where $s$ is the desired sample size and $b$ is the number of burn-in realizations. The last $s$ realizations correspond to the sample of arrangements from the classic Potts model. 
\end{enumerate}

\subsection{Markov Chain Monte Carlo Maximum Likelihood for the Potts model} \label{sec-MCMCMLEPotts}

Let $\boldsymbol{\theta}_0$ be a fixed, arbitrary value of the parameter $\boldsymbol{\theta}$, and $\mathbf{Y}_1, \mathbf{Y}_2, \ldots, \mathbf{Y}_r$ a sample from the distribution parameterized by $\boldsymbol{\theta}_0$. Then, for an observed arrangement $\mathbf{x}^{obs}$, the log-likelihood ratio $\log\left(\frac{L(\mathbf{x}^{obs}, \boldsymbol{\theta})}{L(\mathbf{x}^{obs}, \boldsymbol{\theta}_0)}\right)$ is: 

\begin{equation}\label{eq:loglikratio}
    l(\boldsymbol{\theta})-l(\boldsymbol{\theta}_0)=(\boldsymbol{\theta}-\boldsymbol{\theta}_0)^t\mathbf{G}(\mathbf{x}^{obs})- \log \left(\frac{c(\boldsymbol{\theta})}{c(\boldsymbol{\theta}_0)}\right)
\end{equation}

In equation (\ref{eq:loglikratio}), the second term can be rewritten as 

\begin{equation} \label{eq:ratioconstants}
    \frac{c\boldsymbol{\theta})}{c(\boldsymbol{\theta}_0)}=\frac{1}{c(\boldsymbol{\theta}_0)}\int \exp \{\boldsymbol{\theta}^t\mathbf{x}\}d\mathbf{x}=E_{\boldsymbol{\theta}_0}\left[\exp \{(\boldsymbol{\theta}-\boldsymbol{\theta}_0)^t \mathbf{G}(\mathbf{Y})\}\right]
\end{equation}

By substituting the result from equation (\ref{eq:ratioconstants}) in equation (\ref{eq:loglikratio}), we obtain \\

\begin{equation} \label{eq:expectedvalApp}
    l(\boldsymbol{\theta})-l(\boldsymbol{\theta}_0)=(\boldsymbol{\theta}-\boldsymbol{\theta}_0)^t\mathbf{G}(\mathbf{x}^{obs})-\log \left(E_{\boldsymbol{\theta}_0}\left[\exp \{(\boldsymbol{\theta}-\boldsymbol{\theta}_0)^t \mathbf{G}(\mathbf{Y})\}\right]\right)
\end{equation}

The expected value in equation (\ref{eq:expectedvalApp}) can be approximated using the sample $\mathbf{Y}_1, \mathbf{Y}_2, \ldots, \mathbf{Y}_r$, which can be obtained using the procedure described in Section \ref{sec-PottsSimulation} once the value of $\boldsymbol{\theta}_0$ has been chosen. Then,
\begin{equation}\label{eq:naiveapprox}
    l(\boldsymbol{\theta})-l(\boldsymbol{\theta}_0) \approx (\boldsymbol{\theta}-\boldsymbol{\theta}_0)^t\mathbf{G}(\mathbf{x}^{obs})-\log\left[\frac{1}{r}\sum_{j=1}^{r}\exp\{(\boldsymbol{\theta}-\boldsymbol{\theta}_0)^t\mathbf{G}(\mathbf{Y}_{j})\}\right]
\end{equation}

The function in equation (\ref{eq:naiveapprox}) can be maximized using a numerical optimization method. \cite{Geyer1991} states that the Monte Carlo MLE $\hat{\boldsymbol{\theta}}_{m}$ converges to the true MLE as $r$ goes to infinity. 

\cite{Hummel2012} called the approximation in equation (\ref{eq:naiveapprox}) the naive approximation and presents an application where it degrades rapidly as $\boldsymbol{\theta}_0$ moves away from the maximum likelihood estimate $\hat{\boldsymbol{\theta}}$. They also proposed an alternative approximation for the log-likelihood ratio, based on the assumption that the random variables $\mathbf{Z}_i=(\boldsymbol{\theta}-\boldsymbol{\theta}_0)^t\mathbf{G}(\mathbf{Y}_i)$ follow a normal distribution. Alternatively, \cite{Fellows2012} presents an approximation of the log-likelihood ratio that uses the Taylor expansion of $e^x$ around $0$ and is based on the cumulant generating function of variables $\mathbf{Z}$. The i-th cumulant of $\mathbf{Z}$ is denoted by $\kappa_i$. Then, 

\begin{equation}
\begin{aligned}
\log\left(E_{\boldsymbol{\theta}_0}\left[\exp\left((\boldsymbol{\theta}-\boldsymbol{\theta}_0)^t\mathbf{G}(\mathbf{Y})\right)\right]\right)&=
\log\left(E_{\boldsymbol{\theta}_0}\left[1+(\boldsymbol{\theta}-\boldsymbol{\theta}_0)^t\mathbf{G}(\mathbf{Y})+\frac{\left((\boldsymbol{\theta}-\boldsymbol{\theta}_0)^t\mathbf{G}(\mathbf{Y})\right)^2}{2}+\ldots\right]\right) \\
&=\sum_{i}^{\infty}\frac{\kappa_i}{i!} \approx \sum_{i}^{l}\frac{\hat{\kappa_i}}{i!}  
\end{aligned}
\end{equation}

\cite{Fellows2012} recommends using $2 \leq l \leq 4$ to balance the error of the Taylor expansion and the error from sampling. Let $\mathbf{m}_0$ and $\mathbf{\Sigma}_0$ be the mean vector and the covariance matrix of $\mathbf{G}(\mathbf{Y})$ under the distribution parameterized by $\boldsymbol{\theta}_0$. Then, when $l=2$, the cumulant generating function approximation is equal to

\begin{equation}\label{eq:cum_approx}
      l(\boldsymbol{\theta})-l(\boldsymbol{\theta}_0) \approx \sum_{i}^{2}\frac{\hat{\kappa_i}}{i!} = \sum (\boldsymbol{\theta}-\boldsymbol{\theta}_0)^t\left(\mathbf{G}(\mathbf{x}^{obs})-\hat{\mathbf{m}}_0\right)- \frac{1}{2}(\boldsymbol{\theta}-\boldsymbol{\theta}_0)^t \hat{\mathbf{\Sigma}}_0(\boldsymbol{\theta}-\boldsymbol{\theta}_0),
\end{equation}
and the approximation in Equation (\ref{eq:cum_approx}) is
equal to the log-normal approximation proposed by \cite{Hummel2012} but has no distributional assumptions. The function in Equation (\ref{eq:cum_approx}) is maximized at
\begin{equation}\label{eq:max_cum_approx}
    \boldsymbol{\theta}_0+\hat{\mathbf{\Sigma}}^{-1}_{0} \left(\mathbf{G}(\mathbf{x}^{obs})-\hat{\mathbf{m}}_0\right).
\end{equation}

The cumulant generating function approximation also degrades when $\boldsymbol{\theta}_0$ moves away from $\hat{\boldsymbol{\theta}}$ \citep{Fellows2012}. 

\subsection{Partial stepping algorithm}\label{sec-PartialAlg}

Since the Potts model belongs to the exponential family, the maximum likelihood estimate of $\boldsymbol{\theta}$ is the value $\hat{\boldsymbol{\theta}}$ such that 
\begin{equation}\label{eq:Pottsmle}
    E_{\hat{\boldsymbol{\theta}}}\left[\mathbf{G}(\mathbf{X})\right]=\mathbf{G}(\mathbf{x}^{obs}).
\end{equation}

However, the relationship between $\hat{\boldsymbol{\theta}}$ and $\mathbf{G}(\mathbf{x}^{obs})$ is unknown. The partial stepping algorithm, originally proposed by \cite{Hummel2012}, takes steps toward $\mathbf{G}(\mathbf{x}^{obs})$ by jumping from the mean value parameter space to the parameter space following the next steps:

  \begin{enumerate}
        \item Set $t=0$, and choose some initial parameter vector $\boldsymbol{\theta}_0^{(0)}$.

        \item Use MCMC to simulate samples $\mathbf{Y}_1,\ldots,\mathbf{Y}_r$ from $P_{\boldsymbol{\theta}_0^{(0)}}(\mathbf{Y})$.

        \item Calculate the sample mean $\overline{\boldsymbol{\zeta}}_t=\frac{1}{r}\sum_{i=1}^r \mathbf{G}(\mathbf{y})$.

        \item For some $\gamma_t \in ({0,1}]$, define the pseudo-observation $\widehat{\zeta}_t$ to be
        
        $$ \widehat{\boldsymbol{\zeta}}_t=\gamma_t \mathbf{G}(\mathbf{x}^{obs})+(1-\gamma_t)\overline{\boldsymbol{\zeta}}_t $$
            
We want the value of $\gamma_t$ to be the largest value in the interval $({0,1}]$ such that the observation
\begin{equation*}
    1.05\gamma_t \mathbf{G}(\mathbf{x}^{obs}) + (1-1.05\gamma_t) \widehat{\boldsymbol{\zeta}}_t 
\end{equation*}
is inside the convex hull generated by the value $\mathbf{G}(\mathbf{y}_1),\ldots, \mathbf{G}(\mathbf{y}_r)$.
        \item Replace $\mathbf{G}(\mathbf{x}^{obs})$ for $\widehat{\boldsymbol{\zeta}}_t$ in Equation (\ref{eq:naiveapprox}) and optimize the function if the naive approximation is used. When using the cumulant-generating function approximation, replace $\mathbf{G}(\mathbf{x}^{obs})$ with $\widehat{\boldsymbol{\zeta}}_t$ in Equation (\ref{eq:max_cum_approx}).
        \item The value that maximizes any of the approximations is $\boldsymbol{\theta}_0^{(t+1)}$.
    \end{enumerate}

These steps should be repeated until obtaining $\gamma_t=1$ for two consecutive iterations. The value of $\boldsymbol{\theta}_t$ from the last step of the last iteration will be the parameter that generates the sample $\mathbf{Y}_1, \ldots, \mathbf{Y}_n$ used in Equation (\ref{eq:naiveapprox}) or Equation (\ref{eq:cum_approx}). Optimize the log-likelihood approximation one last time, and the result will be the approximate maximum likelihood estimate of $\boldsymbol{\theta}$, denoted by $\Tilde{\boldsymbol{\theta}}$. 

\begin{figure}
    \centering
    \includegraphics[width=3.2cm, height=6cm]{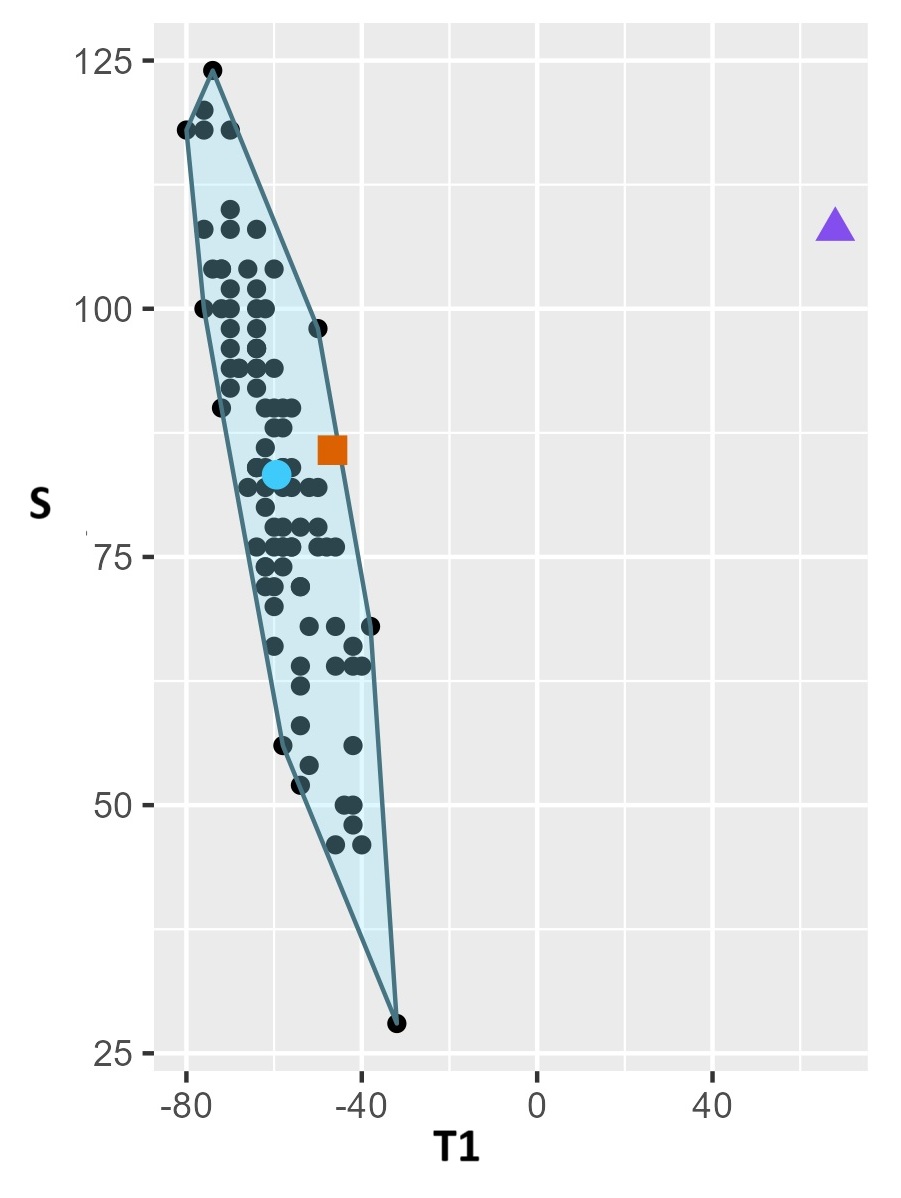}
    \includegraphics[width=3.2cm, height=6cm]{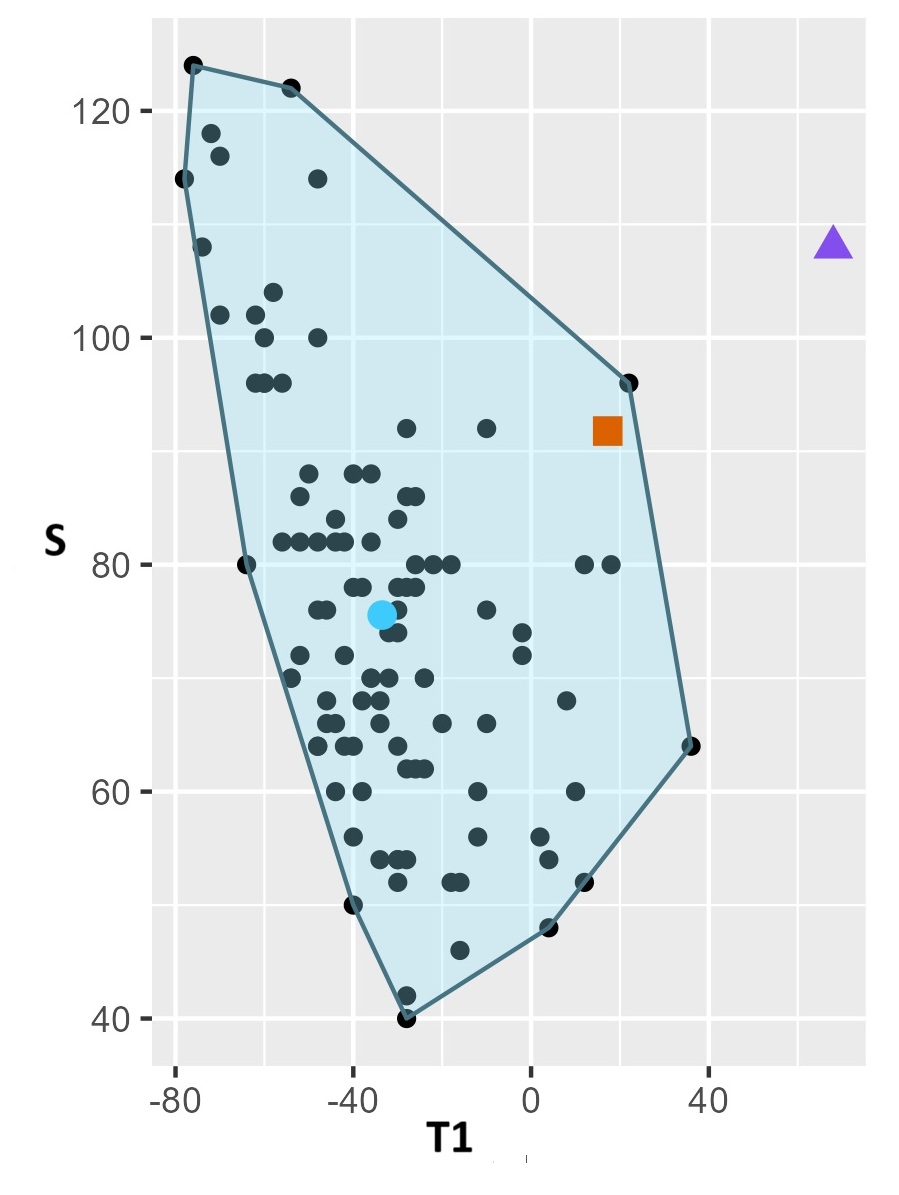}
        \includegraphics[width=3.2cm, height=6cm]{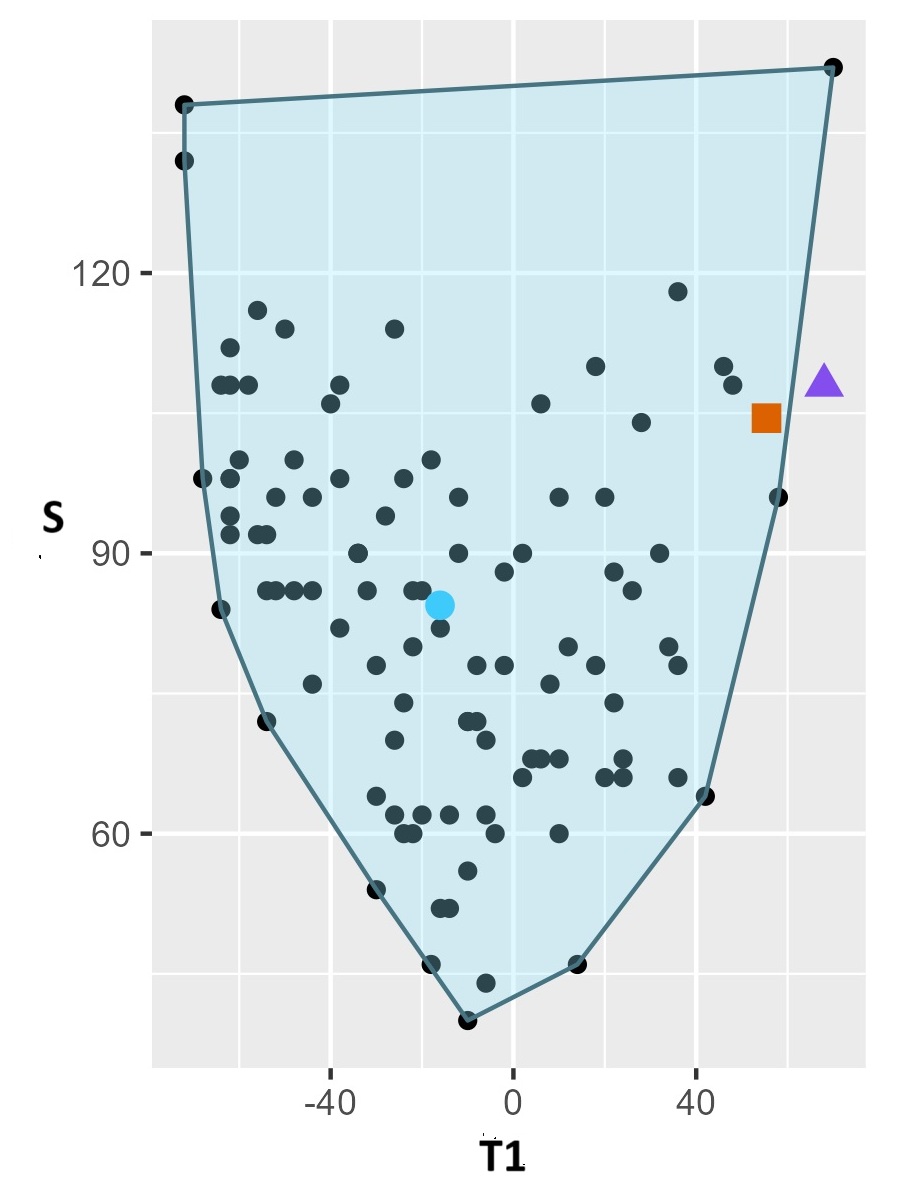}
    \includegraphics[width=4cm, height=6cm]{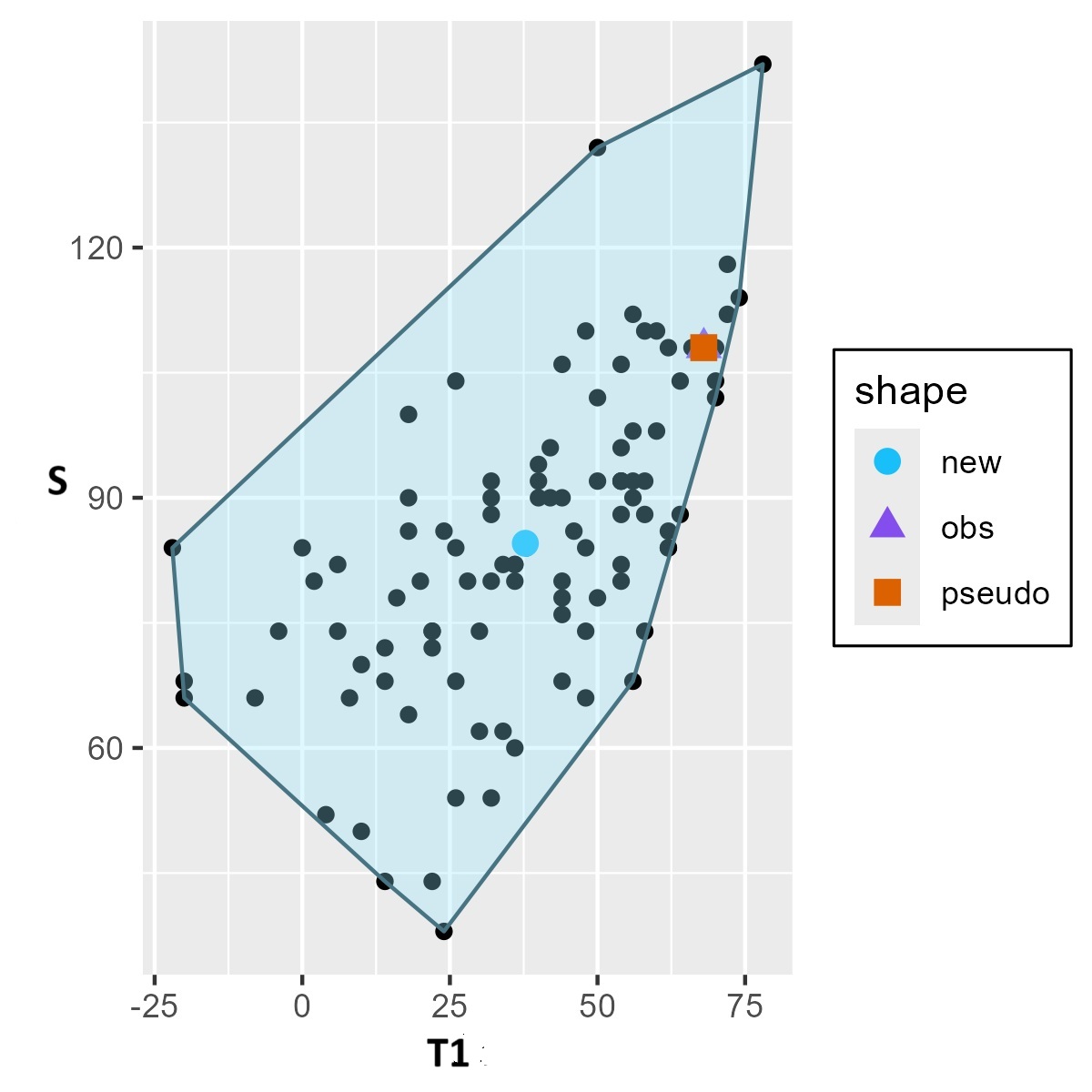}
    \caption{Iterations 1, 2, 3 and 4 in the partial stepping algorithm for the Ising model}
    \label{fig:stepAlgorithm}
\end{figure}

Figure \ref{fig:stepAlgorithm} illustrates the partial stepping algorithm in the particular case of the Ising model. The black dots represent the sample obtained in Step 2 of the algorithm at each iteration $t$, the blue circle represents its sample mean $\overline{\boldsymbol{\zeta}}_t$, the purple triangle represents the original observed value of the sufficient statistics $\mathbf{G}(\mathbf{x}^{obs})$, and the orange square represents the pseudo-observation computed at Step 4. At each iteration, the pseudo-observation is at the edge of the convex hull of the sample in the direction of $\mathbf{G}(\mathbf{x}^{obs})$. In this example, the pseudo-observation reaches the value $\mathbf{G}(\mathbf{x}^{obs})$ in iteration 4, which means that we have found a value of $\boldsymbol{\theta}$ that is close to $\hat{\boldsymbol{\theta}}$ according to the fact in expression \ref{eq:Pottsmle}. Then, we conclude that after 4 iterations, we should run one more iteration and maximize the approximation of the log-likelihood ratio to find an appropriate value of $\boldsymbol{\theta}_0$ for the MCMCMLE procedure.

\subsection{Gibbs sampler the Tapered Potts model}\label{sec-GibbsTapered}

\begin{enumerate}
    \item In step 1, in addition to fixing the values of $\alpha$ and $\beta$, it is also necessary to fix the values of $\boldsymbol{\tau}$ and $\mathbf{m}=(m_1,\ldots,m_{K-1})$. The value of $\boldsymbol{\tau}$ should be fixed according to the recommendations given in the following sections. On the other hand, the value of $\mathbf{m}$ should be the desired center of the distribution of $T(\mathbf{x})$. In this work, we will assume $\mathbf{m}=\mathbf{T}(\mathbf{x}^{obs})$.
        \item Randomly select one value $k$ with $k=1,\ldots,K$. Let $\mathbf{x}$ be an initial value of the observed arrangement for the sampler where all the realizations $x_{i}=k$ for $i=1,\ldots, M$.
    \item Generate a list of neighbors or a contiguity matrix. This is necessary to calculate the change of the statistic $S$ at each iteration. 
    \item For a site $i$, randomly select a value $l$ from the set $\{1,\ldots,k-1, k+1, \ldots, K\}$. Let $\mathbf{z}$ be a new arrangement where $z_{j}=x_{j}$ for $j \neq i$, and $z_{i}=l$. Calculate
    \begin{equation}
    \begin{aligned}
       r&=\text{min}\biggl\{1, \exp \Bigl\{\alpha_l-\alpha_k + \beta \sum_{i \sim j} \left(I\left(l=x_{j}\right) - I\left(x_{i}=x_{j}\right)\right)\\
       &-\sum_{k=1}^{K-1}\tau_k \left(\left(T_{k}(\mathbf{z})-m_k\right)^2-\left(T_{k}(\mathbf{x})-m_k\right)^2\right) \Bigl\} \biggl\}
       \end{aligned}
    \end{equation}
Let $U \sim \text{Uniform}(0,1)$. If $u<r$, then set $x_{i}$. Otherwise, keep the previous value $x_{i}$. Repeat this step for all locations $i=1,\ldots, M$ to complete a full simulation cycle.
    \item For the Tapered model, we will also attempt to do the symmetric swap described in Section \ref{sec-PottsSimulation}. Particularly, generate a new arrangement $\mathbf{y}$ following step 5, (a) from Section \ref{sec-PottsSimulation}. Then, calculate the odds ratio as

$$odds=\exp\Bigl\{ \boldsymbol{\alpha}^{t}\left(\mathbf{T}(\mathbf{y})-\mathbf{T}(\mathbf{x}) \right)- \sum_{k=1}^{K-1}\tau_k \left(\left(T_{k}(\mathbf{y})-m_k\right)^2-\left(T_{k}(\mathbf{x})-m_k\right)^2\right)\Bigl\}$$
If $odds \geq 1$, $\mathbf{y}$ is set as a new observation from the model. Otherwise, $\mathbf{x}$ will be the new element of the sample.
    \item Repeat steps 4 and 5 a total of $s+b$ times, where $s$ is the desired sample size and $b$ is the number of burn-in realizations. The last $s$ realizations correspond to the sample of arrangements from the Tapered Potts model. 
\end{enumerate}

Note that the Tapered Potts model no longer has the Markovian property, and therefore, it is not possible to implement more efficient algorithms for sampling, like Swendsen-Wang.

\subsection{Approximate maximum-likelihood estimates for simulation study}

\begin{table}
    \centering
    \begin{tabular}{|c|c|c|c|c|c|c|}
    \hline
        Scenario & Method & $\hat{\alpha}_1$ & $\hat{\alpha}_2$ & $\hat{\alpha}_3$ &  $\hat{\beta}$ & $\tau$ \\
        \hline
        \multirow{2}{4em}{1}& Pseudolikelihood & -0.04968 & -0.0425 & -0.0298 & 0.6934 & NA \\
       & Maximum likelihood & -0.0497 & -0.0316 & -0.0022 & 0.6046 & NA  \\
        \hline
        \multirow{2}{4em}{2}& Pseudolikelihood & 0.0504 & 0.0082 & -0.0027 & 1.1235  & NA \\
       & Maximum likelihood & 0.0032 & -0.0064 & -0.0056 & 1.1218 & 0.0526 \\
       \hline
        \multirow{2}{4em}{3}& Pseudolikelihood & 0.5371 & 0.3475 & -0.0270 & 0.6308  & NA \\
       & Maximum likelihood & -0.0321 & 0.0506 & 0.0394 & 1.2049 & 0.495 \\
       \hline
          \multirow{2}{4em}{4}& Pseudolikelihood & 0.7041 & 0.5337 & 0.1214 & 0.6092  & NA \\
       & Maximum likelihood & 0.8012 & 0.5550 & 0.1209 & 0.5302 & NA \\
       \hline
        \multirow{2}{4em}{5}& Pseudolikelihood & 0.1473 & 0.1214 & -0.0549 & 1.0446  & NA \\
       & Maximum likelihood & 0.0356 & 0.0031 & -0.0375 & 1.0398 & 0.000198 \\
       \hline
       \multirow{2}{4em}{6}& Pseudolikelihood & -0.0725 & 0.0367 & 0.1462 & 1.7244  & NA \\
       & Maximum likelihood & -0.0165 & -0.0130 & -0.0222 & 1.1630 & 0.001443 \\
       \hline

    \end{tabular}
    \caption{Parameter estimates from pseudolikelihood and maximum-likelihood inference in all simulation scenarios}
    \label{tab:MLE_results_Sim}
    \end{table}

    Analyzing the results of Table \ref{tab:MLE_results_Sim}, we see that pseudo-likelihood estimates are fairly similar to the maximum-likelihood estimates in scenarios 1 and 4. However, they start to differ as the spatial correlation increases. This is explained by the lack of fit that the Potts model presents in the presence of high correlation. We also highlight that the ideal value of $\tau$ in scenarios 2 and 3 is relatively large, which means that the variability of statistics $T_{k}(\mathbf{x})$ is being highly constrained and the bimodality coefficient criteria to find the ideal value of $\tau$ may be too strict in these cases.
\end{document}